\documentclass[twocolumn]{aastex631}

\usepackage{enumitem}
\usepackage{lipsum}  
\usepackage{graphicx}
\usepackage{color}
\usepackage{amsmath}
\usepackage{multirow}
\usepackage{booktabs}
\usepackage{xspace}
\definecolor{PineGreen}{rgb}{0., 0.62,0.25}

\newcommand{\Gaia}{{\textit{Gaia}}\xspace}
\newcommand{\vesc}{v_{\rm{esc}}}
\newcommand{\vmin}{v_{\rm{min}}}

\renewcommand{\arraystretch}{1.8}

\begin{document}

\title{The Escape Velocity Profile of the Milky Way from \textit{Gaia} DR3}



\author[0000-0002-3400-6991]{Cian Roche}
\affiliation{Department of Physics and MIT Kavli Institute for Astrophysics and Space Research, \\
77 Massachusetts Avenue, Cambridge, MA 02139, USA}

\author[0000-0003-2806-1414]{Lina~Necib}
\affiliation{Department of Physics and MIT Kavli Institute for Astrophysics and Space Research, \\
77 Massachusetts Avenue, Cambridge, MA 02139, USA}
\affiliation{The NSF AI Institute for Artificial Intelligence and Fundamental Interactions, \\
77 Massachusetts Avenue, Cambridge, MA 02139, USA}

\author[0000-0003-4969-3285]{Tongyan Lin}
\affiliation{Department of Physics, University of California San Diego, La Jolla, CA 92093, USA}

\author[0000-0002-4669-9967]{Xiaowei Ou}
\affiliation{Department of Physics and MIT Kavli Institute for Astrophysics and Space Research, \\
77 Massachusetts Avenue, Cambridge, MA 02139, USA}

\author[0000-0001-6189-8457]{Tri Nguyen}
\affiliation{Department of Physics and MIT Kavli Institute for Astrophysics and Space Research, \\
77 Massachusetts Avenue, Cambridge, MA 02139, USA}
\affiliation{The NSF AI Institute for Artificial Intelligence and Fundamental Interactions, \\
77 Massachusetts Avenue, Cambridge, MA 02139, USA}


\begin{abstract}
The escape velocity profile of the Milky Way offers a crucial and independent measurement 
of its underlying mass distribution and dark matter properties. 
Using a sample of stars from the third data release of \Gaia~with 6D kinematics and strict quality cuts, we obtain an escape velocity profile of the Milky Way from $4\,\rm{kpc}$ to $11\,\rm{kpc}$ in Galactocentric radius. To infer the escape velocity in radial bins, we model the tail of the stellar speed distribution with both traditional power law models and a new functional form that we introduce. 
While power law models tend to rely on extrapolation to high speeds, we find our new functional form gives the most faithful representation of the observed distribution. Using this for the escape velocity profile, we constrain the properties of the Milky Way's dark matter halo modeled as a Navarro-Frenck-White profile. Combined with constraints from the circular velocity at the solar position, we obtain a concentration and mass of $c_{200\rm{c}}^{\rm{DM}} = 13.9^{+6.2}_{-4.3}$ and $\rm{M}_{200\rm{c}}^{\rm{DM}} = 0.55^{+0.15}_{-0.14}\times 10^{12} \, M_\odot$. 
This corresponds to a total Milky Way mass of $\rm{M}_{200\rm{c}} = 0.64^{+0.15}_{-0.14}\times 10^{12} \, M_\odot$, which is on the low end of the historic range of the Galaxy's mass, but in line with other recent estimates. 
\end{abstract}


\keywords{\href{http://astrothesaurus.org/uat/1051}{Milky Way dynamics (1051)}; \href{http://astrothesaurus.org/uat/1608}{Stellar kinematics (1608)}; \href{http://astrothesaurus.org/uat/598}{Galaxy stellar halos (598)}}


\section{Introduction} \label{sec:intro}
The earliest hints of the existence of Dark Matter (DM) came from the dynamics of stars and galaxies \citep{zwicky:1933,rubin:1970}. 
These observations provided the first evidence that galaxies are surrounded by DM halos extending more than an order of magnitude past the visible matter \citep{1974ApJ...193L...1O,1974Natur.250..309E}. In the Milky Way, the DM halo dominates the mass of the Galaxy, and is expected to be $\sim$ 1-2 orders of magnitudes larger in mass than the baryonic (stars and gas) component. However, many of its detailed properties have thus far not been well understood. 
Insight into both the total mass and density profile of the DM halo is extremely valuable for galactic dynamics and studies of DM, informing our understanding of baryonic feedback \citep{2016MNRAS.456.3542T,2020MNRAS.497.2393L}, the particle nature of DM \citep{Spergel:1999mh,Tulin:2017ara,Lisanti:2018qam,Nadler:2019zrb,DES:2020fxi}, and indirect detection of DM \citep{Cirelli:2010xx,Slatyer:2017sev,Rinchiuso:2020skh}.

Many methods have been used to measure the MW mass, and to a lesser extent the density profile of DM: the circular velocity of the MW \citep{nesti:2013, deSalas:2019, eilers:2019,Ou:2023,2023arXiv230900048J,2023ApJ...942...12W,2023ApJ...946...73Z}, the population and motion of satellites \citep{barber:2014, cautun:2014, patel:2018, callingham:2019, sohn:2018, watkins:2019, fritz:2020}, stellar streams \citep{gibbons:2014, kupper:2015, dierickx:2017, Vasiliev:2021}, and the escape velocity of the MW \citep{Piffl2014, williams:2017, banik:2018, Monari2018, deason:2019, Koppelman2021, Necib2022, Prudil:2022}. The escape velocity, in particular, gives a measure of the gravitational potential at a given location, and can be used to constrain the DM potential when combined with models of the baryonic mass. It is typically extracted by modeling the tail of the stellar speed distribution. Due to a lack of data, earlier studies focused on the local escape velocity in the solar neighborhood \citep{Leonard1990,smith:2007}. 
In this paper, we produce precise measurements of the escape velocity over a range of Galactocentric radii, in order to obtain an escape velocity profile and thus inform the DM potential over a wider range. We revisit prior power law models for the tail of the speed distribution, and also introduce a new, more robust functional form.

Our analysis is made possible with the significant improvements in the third data release from \Gaia~(\Gaia~DR3). The \Gaia space mission \citep{gaia:2016, gaiaDR1, gaiaDR2, GaiaeDR3} has revolutionized the field of MW astronomy, from studies of streams \citep{malhan:2018, price-whelan:2018, helmi:2020, chandra:2023}, dust \citep{green:2019, lallement:2019, leike:2020}, disk resonances \citep{kawata:2018, trick:2021}, to the discovery of previously unknown merging events \citep{2018Natur.563...85H,2018MNRAS.478..611B,2020Necib,Naidu2020}, and more. The latest data release of the \Gaia mission includes positions, parallax, proper motions, and radial velocities of an astounding 33~million stars \citep{gaiaDR3, gaiaRVs}. The six-dimensional features of this large dataset allow us to finally probe detailed dynamics of the MW at a few~kpc from the solar position, enabling more precise and reliable determinations of the escape velocity profile. \Gaia~DR3, in particular, has a significant advantage over DR2 in that it includes many more sources at large distances from the Sun with line-of-sight velocities. 

Even with excellent data, estimating the escape velocity crucially relies on modeling the tail of the stellar speed distribution.  \cite{Leonard1990} introduced a power-law model, with a distribution given by
\begin{equation} \label{eq:powerlaw}
    g(v \mid \vesc, k) \propto (\vesc - v)^{k}, \quad v\in[\vmin,\vesc]
\end{equation}
where $\vesc$ is the escape velocity, $k$ is the slope of the power law, and $\vmin$ is a lower speed cutoff which reflects the fact that we only model the \textit{tail} of the speed distribution for stars at a given Galactocentric radius. The overall normalization is set such that $\int_{\vmin}^{\vesc} dv \, g(v) = 1$. This functional form has been the basis of most of the escape velocity measurements in the MW for the last few decades. However, escape velocity estimates based upon this modeling \citep{smith:2007, Piffl2014, Monari2018, deason:2019, Koppelman2021, Necib2022} face various challenges such as small sample sizes, degeneracy of parameters, systematic uncertainties in model selection, and the choice of strong prior distributions on the fitting parameters. 

The functional form  in Eq.~\ref{eq:powerlaw} exhibits a strong degeneracy between the two parameters $\vesc$ and $k$, leading to large uncertainties on the escape velocity estimates; this degeneracy can be exacerbated by small sample sizes or if the underlying distribution deviates from a single power law. To reduce such uncertainties on $\vesc$, previous work in the literature adopted an array of either theoretically or numerically (based on cosmological simulations) motivated priors on the slope $k$. Such choices affect the final measurement of the escape velocity, and ideally one would require a model that best fits the data without the need to adopt strong priors. 

\begin{figure*}[t!]
  \centering
  \includegraphics[width=\textwidth]{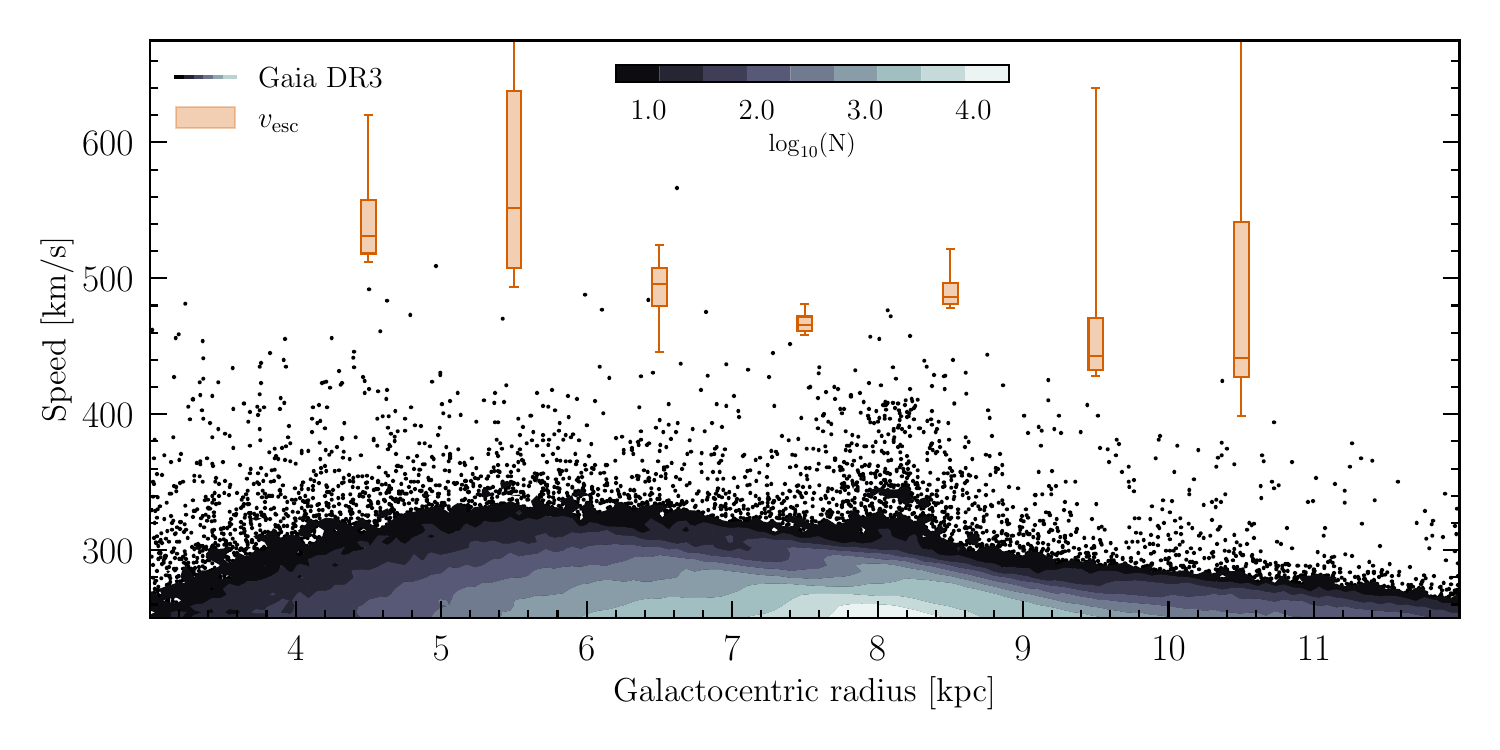}
  \caption{High-speed and high-quality sample of stars in \Gaia~DR3 with 6D kinematics, and the derived escape velocity profile at $68\%$ (boxes) and $95\%$ (whiskers) confidence intervals. Medians marked with a horizontal line. Escape velocities shown here are obtained by binning the data into $1\,\rm{kpc}$ bins and fitting the speed distributions above $\vmin = 310\,\rm{km\,s^{-1}}$ with the stretched exponential power law model introduced in this paper. Contours correspond to the logarithm of the number of stars at that speed and Galactocentric radius, labelled N.}
  \label{fig:data}
\end{figure*}

Motivated by the discovery of the \Gaia-Sausage-Enceladus (GSE) \citep{2018Natur.563...85H,2018MNRAS.478..611B} merger that has contributed a large fraction of the ex-situ stars in the MW \citep[see e.g.][]{Necib:2019a,naidu:2020}, and arguing for the need to address the issues above, \cite{Necib2022methods} adopted a two-component power law, generalizing Eq.~\ref{eq:powerlaw} but with two different slopes $k_1$ and $k_2$, 
\begin{align} \label{eq:powerlaw2}
    g(v \mid \vesc, k_1, k_2) \propto &(1-f_S) g(v \mid \vesc, k_1)   \\ 
    & + f_S g(v \mid \vesc, k_2), \quad v\in[\vmin,\vesc] \nonumber
\end{align}
where $f_S$ is an additional parameter that controls the relative fraction of the power laws. Based on this new functional form, \cite{Necib2022} computed the escape velocity of the MW at the position of the Sun using the early \Gaia data release 3 (eDR3) \citep{GaiaeDR3}. 

The two-component model in \cite{Necib2022methods} provided a more general functional form, allowing for a better fit to data containing substructure. Indeed, \cite{Necib2022} showed using the Akaike information criterion (AIC; \cite{AIC}), that the two-component model is preferred at lower $\vmin$, but that at higher $\vmin$ it was not distinguishable from a single component. This affirms the presence of a deviation from the single power law away from $\vesc$. The model further addressed the degeneracy and the issue of strong priors on the slope $k$: with the modified distribution, it was possible to obtain a good fit to the data and robust measurement of $\vesc$ with wide priors on the slopes.

In this work, we aim at further investigating models for the tail of the speed distribution, adapting to the improved quality and increased statistics of the data in \Gaia DR3.  We reconsider the single power law and  two-component power law models, and obtain the first two-component power law escape velocity profile using this extended dataset. With the improved dataset, we find that there remains a number of issues in using these power-law models: they can predict a high $\vesc$ relative to the fastest stars in the dataset, results can be sensitive to the choice of $\vmin$, and some parameters correlated with the escape velocity can remain unconverged even with wide priors. Motivated by these challenges, we introduce a new functional form that allows for a steeper rise in the speed distribution away from $\vesc$, while keeping a power law cutoff near $\vesc$. We find that this new form most closely tracks the fastest stars and gives the most stable results. With each of these models, we obtain independent escape velocity measurements in 1~kpc-wide bins from 4~kpc to 11~kpc in Galactocentric radius. We use the resulting escape velocity profile to fit the DM density profile, and combine this with a circular velocity measurement from \cite{eilers:2019} to obtain an updated measurement of the MW mass.

The remainder of this paper is organized as follows: In Section \ref{sec:data}, we present the selection criteria for the high-quality sample of \Gaia~DR3 data used to obtain our escape velocity profile and Milky Way halo constraints. In Section \ref{sec:methods} we describe the general modeling approach for the tail of the stellar speed distribution, and in Section \ref{sec:modeling} we present the different models for this speed distribution. These models include the previously studied one and two-component power laws, as well as the new model we introduce in this work, the ``stretched exponential power law." In Section \ref{sec:results}, we show the results of applying these three models to the \Gaia~DR3 data, presenting the escape velocity as a function of the radial distance and validating the modeling. Finally in Section \ref{sec:DM}, we use the escape velocity data to constrain the DM density profile and total mass of the Milky Way.

\section{Data} \label{sec:data}

To evaluate the escape velocity of the Milky Way, we use the third catalog from the \Gaia space mission (\Gaia~DR3) \citep{gaia:2016,gaiaDR3}, which contains $\sim 33$ million stars with 6D kinematics, an increase of about a factor of five from the previous release of \Gaia~eDR3 \citep{GaiaeDR3}. 
As we are interested in modeling the tail of the speed distribution, we extract a high-quality sample of stars from this catalog in order to avoid stars with contaminated radial velocities and/or unphysical parallax measurements. Such contaminants could bias escape velocity estimates toward higher values since these analyses are sensitive to the fastest star in the sample (see e.g. \cite{Koppelman2021}). 

To obtain a high quality sample, we follow the selection criteria of \cite{marchetti}, imposing quality cuts on parallax measurements, renormalized unit weight error (RUWE), number of transits used to calculate the radial velocity of the source, and signal-to-noise ratio. We do however demand  higher-quality signal to noise ratios of $10$ (as opposed to $5$) in line with \cite{gaiaRVs}, which found a large number of spurious radial velocity measurements near the Solar position at low signal-to-noise, and parallaxes with at most $10\%$ error (as opposed to $20\%$) to reduce the contamination of stars across radial bins. 
For parallax measurements, we apply a zero-point correction according to \cite{zpcorr} and choose not to include those stars for which a zero-point-corrected and positive parallax is unavailable. Note that an alternative implementation of the cut on parallax percentage errors would be a ``bin membership" cut whereby a star only be included in the analysis if it is a member of its hosting radial bin at some chosen confidence level. In the present analysis, such a cut would be too restrictive and reduce statistics away from the solar position due to the limitations of parallax measurements. This may lead to some leakage of stars near bin edges across radial bins, but since the escape velocity is not expected to vary significantly over distances $\lesssim 0.5\,\rm{kpc}$, we do not expect this to appreciably bias results as is shown in Fig. \ref{fig:offsetbins}, in which we repeated our analysis with a shift in the binning. 

We also require that the derived percentage error on the speed be less than $1\%$ since large errors can significantly modify the shape of the tail of the distribution. In some works such as \cite{Piffl2014, Monari2018}, only the retrograde stars in the sample are modeled to avoid disk contamination, but following \cite{Necib2022} we can also remove disk stars via high minimum speed cuts, here not considering stars with speeds below $300\,\rm{km\,s^{-1}}$, or by choosing more flexible models that can account for the presence of additional velocity components.

The cuts are summarized as follows:
\setlist{nolistsep}
\begin{itemize}[itemsep=0em, label=-]
    \item \texttt{RUWE} $\leq 1.4$
    \item \texttt{RV\_NB\_TRANSITS} $\geq10$   
    \item \texttt{RV\_EXPECTED\_SIG\_TO\_NOISE} $\geq10$   
    \item Positive zero-point-corrected parallax $\varpi - \varpi_{zp} > 0$
    \item Parallax error: $\sigma_{\varpi}/(\varpi - \varpi_{zp}) \leq 10\%$
    \item Minimum speed: $v\geq 300\,\rm{km\,s^{-1}}$ 
    \item Speed error: $\sigma_v/v \leq 1\%$
\end{itemize}

The data processing pipeline\footnote{Pipeline and data products are available at \url{https://doi.org/10.5281/zenodo.8088365}.} used to calculate the stellar kinematics and associated uncertainties in Galactocentric coordinates is similar to that of \cite{Necib2022}, with the addition of the zero-point correction of parallaxes and stricter quality cuts. In particular, we use sources with parallax errors of $10\%$ or less (rather than $20\%$) to reflect the finer radial binning used in this analysis ($1\,\rm{kpc}$ bins as opposed to $2\,\rm{kpc}$), and percentage speed errors of at most $1\%$ (rather than $5\%$) because speed measurement errors can dominate the shape of the high-speed tail; these choices are enabled by the improved statistics of \Gaia~DR3. The Solar position used in this analysis is $x_\odot = -8.122\,\rm{kpc}$ \citep{gravity:2018}, $y_\odot = 0\,\rm{kpc}$, and $z_\odot=0.0208\,\rm{kpc}$ \citep{Bennett:2018} in Galactocentric coordinates. For the Solar peculiar velocity vector we use $v_\odot = (12.9, 245.6, 7.78)\,\rm{km\,s^{-1}}$ \citep{Drimmel:2018}. 

The cuts we utilize result in one of the highest-quality 6D kinematics datasets used for the escape velocity measurements to date. It is also the largest, containing $12,317$ stars above a speed of $300\,\rm{km\,s^{-1}}$, compared to the $3,932$ within $1\,\rm{kpc}$ of the solar position of \cite{Necib2022}, and the $2,067$ in the 6D kinematics sample of \cite{Koppelman2021}. We represent this selection of stars in \Gaia~DR3 in Figure \ref{fig:data}, along with one of the escape velocity profiles obtained in this paper, discussed in detail in Sec. \ref{sec:results}. 

In Figure \ref{fig:data}, one can note some general features of the data. First, there is a slight overdensity of high velocity stars near the solar position. This excess may be due to a larger number of spurious radial velocity measurements in this region \citep{gaiaRVs}, as this feature appears primarily close to the solar position. 

Second, the speed of the fastest stars generally decreases from $\sim 4.5$ to $12\,\rm{kpc}$, partly as a result of the increasing Galactic potential. At Galactocentric radii smaller than $4\,\rm{kpc}$, however, there is a decrease in statistics resulting from parallax quality cuts, leading to fewer high speed stars. In general, the number of observed stars in each radial bin is modulated by the \Gaia~selection function, which is typically modeled as a function of magnitude and sky position \citep{gaiaeDR3selection}, or a combined metric of both parameters such as the median magnitude of a patch of sky \citep{gaiaDR3selection}. 
Assuming no significant correlations between completeness and speed, we  expect the shape of the speed distribution in a given radial bin is not strongly biased by the selection function. However, even if the selection function impacts the detailed shape of the distribution, our goal is to extract the cutoff of the speed distribution and not its exact shape.

One can also note over and under-densities in the contours from $7-8\,\rm{kpc}$ at speeds close to and below $300\,\rm{km\,s^{-1}}$. The source of this feature is the resonant band structure of the speed distribution of the Milky Way \citep{Antoja2018}, and in fact there are other parallel bands which are simply not visible in this limited data sample. Over and under-densities at certain Galactocentric radii cannot be modeled by simple power law distributions, and thus fits to single power laws could lead to biased escape velocity estimates.
This already suggests the need for models that are robust to the presence of such features.

\section{Methods} \label{sec:methods}
In this Section, we present our approach to modeling the tail of the stellar speed distribution, the likelihood for a given set of observed data (speeds of stars in one bin of Galactocentric radius), and the method for obtaining posterior distributions of the model parameters.  The speed distribution will be modeled as the combination of one or more ``bound component(s)" corresponding to kinematic structures that are gravitationally bound to the Galaxy, and an outlier distribution which models both unbound and potentially mismeasured stars. Here we focus on the methodology, while Section~\ref{sec:modeling} will discuss the different parametric models including the single power law Eq. \ref{eq:powerlaw}, the 2-component power law  Eq. \ref{eq:powerlaw2} and the newly introduced ``stretched exponential power law."

\subsection{Bound component}
We first describe the parametric modeling of a single bound component, i.e. an approximately isotropic distribution of stars bound to the Milky Way. 
Let us assume that in a given bin of Galactocentric radius, the tail of the speed distribution for these bound stars is described by $g(v \mid \mathbf{\theta})$ (see Eq.~\ref{eq:powerlaw}), where $\mathbf{\theta}$ is a vector of model parameters including the escape velocity $\vesc$, and $g$ is defined on $[0,v_{\mathrm{esc}}]$ as we expect the distribution of bound stars to be zero at speeds higher than $\vesc$ \citep{Leonard1990}. In the next section, we will consider different functional forms for $g$ with various motivations. 

In \cite{Koppelman2021} and \cite{Necib2022methods}, the probability of observing a star labelled $\alpha$ with a speed $v_\alpha$ and a Gaussian measurement uncertainty $\sigma_\alpha$ is given by the convolution 
\begin{equation}\label{eq:convolution}
    p_\alpha\left(v_\alpha \mid \mathbf{\theta}\right) = C_\alpha(\mathbf{\theta}) \int_0^{v_{\mathrm{esc}}} d v \,\frac{e^{-\frac{\left(v-v_\alpha\right)^2}{2 \sigma_\alpha^2}}}{\sqrt{2 \pi \sigma_\alpha^2}}\,g(v \mid \mathbf{\theta})
\end{equation}
where $C_\alpha(\theta)$ is a normalization constant obtained by integrating over the \textit{data region} $[v_{\mathrm{min}},\infty]$ such that
\begin{equation}\label{eq:normalizeprob}
    \int_{v_\mathrm{min}}^{\infty} d v_\alpha\,p_\alpha\left(v_\alpha \mid \mathbf{\theta}\right) = 1\,.
\end{equation}
However, in this work we utilize a strict quality cut on stellar speed uncertainties $\sigma_{\alpha} / v_\alpha \leq 1\%$. In the limit of vanishing relative uncertainty, the probability distribution Eq. \ref{eq:convolution} reduces to 
\begin{equation}\label{eq:low_uncertainty}
p_\alpha (v_\alpha \mid \theta) = g(v_\alpha \mid \mathbf{\theta})
\end{equation}
with the appropriate normalization according to Eq. \ref{eq:normalizeprob}, where the upper limit becomes $\vesc$ since $g$ has support on $[0,v_{\mathrm{esc}}]$. We utilize the approximation Eq. \ref{eq:low_uncertainty} in the present analysis.

\subsection{Outliers}
Following \cite{williams:2017} and \cite{Necib2022methods}, we include a wide Gaussian outlier distribution, corresponding to unbound stars that are not modeled by the distribution $g(v \mid \mathbf{\theta})$ or stars with potentially mismeasured speeds. The probability for a star $\alpha$ to be drawn from this outlier distribution is modeled as 
\begin{equation}\label{eq:outlier}
    p_\alpha^{\mathrm{out}}\left(v_\alpha \mid \sigma_{\mathrm{out}}\right)=A \exp \left(-\frac{v_\alpha^2}{2\left[\sigma_{\mathrm{out}}^2+\sigma_{ \alpha}^2\right]}\right)
\end{equation}
where the dispersion of the outlier distribution $\sigma_{\mathrm{out}}$ is treated as an additional model parameter. We add the measurement uncertainty in quadrature to the dispersion of the underlying outlier distribution as these quantities are independent, however $\sigma_\alpha\ll \sigma_{\mathrm{out}}$  so this addition has little impact. Furthermore, due to the high quality cuts imposed in Sec. \ref{sec:data}, we expect a small degree of outlier contamination in our sample. We again normalize this distribution over the \textit{data region} $[v_\mathrm{min},\infty]$, finding 
\begin{equation}
    A^{-1}=\sqrt{\frac{\pi}{2}\left(\sigma_{\mathrm{out}}^2+\sigma_{\alpha}^2\right)} \,\mathrm{erfc}\left(\frac{v_{\min }}{ \sqrt{2\left(\sigma_{\mathrm{out}}^2+\sigma_{\alpha}^2\right)}}\right) .
\end{equation}

\subsection{Parameter estimation}
For a bound kinematic structure modeled by the distribution $g(v \mid \mathbf{\theta})$ and an outlier component modeled by Eq. \ref{eq:outlier}, the likelihood for a star $\alpha$ with speed $v_\alpha$ and measurement uncertainty $\sigma_\alpha$ to be drawn from the combined distribution is 
\begin{equation}
    \mathcal{L}_\alpha=(1-f_\mathrm{out})\, p_\alpha\left(v_\alpha \mid \mathbf{\theta}\right)+f_\mathrm{out} \,p_\alpha^{\mathrm{out}}\left(v_\alpha \mid \sigma_{\mathrm{out}}\right)
\end{equation}
where $f_\mathrm{out}$ is the fraction of the integrated distribution contributed by the outlier model, and the probability $p_\alpha$ implicitly depends on the model choice $g(v \mid \mathbf{\theta})$. Had we instead wished to model a combination of $Q$ different bound components each with speed distribution $g_i(v \mid \mathbf{\theta}_i)$ and a parameter $f_i$ describing the fractional contribution of that component to the total distribution, then the likelihood for a single star would take the form
\begin{equation}\label{eq:multiple_components}
    \mathcal{L}_\alpha^{Q}=(1-f_\mathrm{out})\, \left[\sum_{i =1}^Q f_i\, p_\alpha^i\left(v_\alpha \mid \mathbf{\theta}_i \right)\right]+f_\mathrm{out} \,p_\alpha^{\mathrm{out}}\left(v_\alpha \mid \sigma_{\mathrm{out}}\right)
\end{equation}
where $p_\alpha^i$ denotes the probability distribution in Eq. \ref{eq:convolution}, reducing to $g_i(v \mid \mathbf{\theta}_i)$ in the limit of vanishing relative speed uncertainties. In this notation, the 2 and 3-component power law models of \cite{Necib2022methods,Necib2022} would have $Q = 2$ or $3$ and each $g_i$ described by the single power law Eq. \ref{eq:powerlaw} with unique slope $k_i$ but a common escape velocity $\vesc$.

For a set of $N$ stars observed with speeds and Gaussian uncertainties on those speeds $\{(v_\alpha, \sigma_\alpha)\}_{\alpha = 1,\dots,N}$, and a chosen set of model parameters $\mathbf{\theta}$ (or $\{\theta_i\}_{i\in 1,\dots ,Q}$), the log-likelihood would be
\begin{equation}
    \log \mathcal{L} = \sum_{\alpha=1}^N \log \mathcal{L}_\alpha.
\end{equation}
The goal of this analysis is to estimate the posterior distribution of the parameters $\sigma_{\rm{out}}, f_{\rm{out}}$, and $\mathbf{\theta}$ (or $\{\theta_i\}_{i\in 1\dots Q}$), focusing in particular on the parameter $\vesc$. In practice we utilize an implementation\footnote{Code is provided at \url{https://github.com/CianMRoche/MCJulia.jl}, and is a modified version of an earlier implementation available at \url{https://github.com/mktranstrum/MCJulia.jl}.} of the affine-invariant Markov chain Monte Carlo algorithm of \cite{AIMCMC} since it is more efficient in skewed parameter spaces, and work in the \texttt{julia} programming language \citep{julia}, primarily to reduce computational time.

\subsection{Model selection}
The above parameter estimation procedure will be performed for different models of the bound component(s), each detailed in the next section, and we will compare the goodness of fit of these models in part via the Akaike information criterion (AIC; \cite{AIC}), defined as
\begin{equation}\label{eq:AIC}
    \rm{AIC} = 2s - 2\log(\mathcal{L}_{\rm{max}})
\end{equation}
where $s$ is the number of model parameters (for one bound component, it is the number of elements in the vector $\mathbf{\theta}$ plus an additional 2 outlier parameters) and $\log(\mathcal{L}_{\rm{max}})$ is the maximum of the log-likelihood of the model for a given set of data. To compare models $A$ and $B$ one may calculate the $\rm{AIC}$ for each model. Then $\Delta\rm{AIC} \equiv \rm{AIC}_B - \rm{AIC}_A$ positive implies model $A$ is preferred, and conversely $\Delta\rm{AIC}$ negative implies model $B$ is preferred.

\section{Modeling the Tail}\label{sec:modeling}

We now examine the different models $g(v \mid \mathbf{\theta})$ that have been used in the literature to describe bound kinematic components in the tail of the local speed distribution. We summarize the limitations and prior choices of the single power law model in Sec.~\ref{sec:1pl}, then turn to recent developments with multiple power law components in Sec.~\ref{sec:mpl}. These ameliorate some of the challenges in using single component power laws, but still have remaining limitations, as discussed in Sec.~\ref{sec:limitations_2pl}. We then introduce an alternative more robust functional form which addresses these issues in Sec.~\ref{sec:sepl}. We summarize the priors used for all models in this paper in Sec.~\ref{sec:priors}.

\subsection{Single power law (1PL)}
\label{sec:1pl}

The model proposed by \cite{Leonard1990} is that of a single power law,\footnote{Note that when originally proposed, this model was not used with an outlier distribution.} such that the tail of the speed distribution of a collection of stars at similar Galactocentric radius follows 
\begin{equation}\label{eq:singlePL}
    g(v \mid v_\mathrm{esc}, k) \propto (v_\mathrm{esc} - v)^{k},
\end{equation}
as in Eq.~\ref{eq:powerlaw}. This distribution can be understood in the following way: Assuming that Jeans theorem \citep{jeans:1915} holds, and that the velocities are isotropic, the asymptotic distribution function of the speeds of the stars is $g(v \mid \epsilon, k) \propto \epsilon^{k}$ for some index $k$ where $\epsilon = -(\Phi + v^2/2)$ is the total (potential + kinetic) energy \citep{kochanek:1996,smith:2007}. We can identify $\Phi$ with $-\vesc^2/2$ yielding 
\begin{equation}\label{eq:singlePLsquare}
    g(v \mid v_\mathrm{esc}, k) \propto (v_\mathrm{esc}^2 - v^2)^{k}.
\end{equation}
The behavior of this distribution for $v$ close to $\vesc$ is then approximated by Eq.~\ref{eq:singlePL}. Therefore, this single power law model of \cite{Leonard1990} can be understood as the $v\rightarrow \vesc$ limit of the speed distribution of an isotropic system for which Jeans' theorem holds. 
Eq.~\ref{eq:singlePL} is the distribution if both radial (line-of-sight) and tangential (on-sky) velocities are known. If only radial velocities are used, the power law is modified to $k_r = k+1$, an approach taken in \cite{Piffl2014}. If only tangential velocities are known, then $k_t = k + \frac{1}{2}$ \citep{Monari2018,Koppelman2021}.

For analyses using the power-law model, $\vesc$ is highly correlated with the slope parameter $k$, leading to large uncertainties on the resulting escape velocity estimates. To avoid this, priors of various ranges and shapes have been placed on $k$ with various motivations, which we now summarize briefly.

\cite{Leonard1990} argues for $k\in[1,2]$, with $k=1$ for a collisionally relaxed system and $k=1.5$ for an isolated system that has undergone violent relaxation, and \cite{kochanek:1996} adopts $k\in[0.5,2.5]$ to widen the previously proposed prior range. There have also been many prior ranges motivated by different suites of cosmological simulations: \cite{smith:2007} adopts $k\in[2.7,4.7]$ based on simulations outlined in \cite{abadi:2006}, \cite{Piffl2014} and \cite{Monari2018} argue for $k\in[2.3,3.7]$ based on the Aquarius suite \citep{aquarius:2008, scannapieco:2009}, and \cite{deason:2019} investigates the signature of GSE-like mergers in the Auriga suite \citep{auriga:2017} and concludes that an appropriate range is $k\in[1,2.5]$.

Furthermore, the shape of the prior distribution for both $\vesc$ and $k$ varies across these works. For example a uniform prior in $\log(\vesc)$ is often (but not always) chosen to favor lower values of $\vesc$, and \cite{smith:2007} adopts priors in $\vesc$ and $k$ derived from Jeffreys’ rules \citep{jeffrey:1961}. In \cite{Koppelman2021} a local value of $k$ is obtained and its posterior treated as a prior distribution for determining $\vesc$ at other Galactocentric radii. Since the escape velocity is highly correlated with the $k$ parameter, the chosen prior distribution can strongly influence the escape velocity estimate.

\begin{figure*}[ht]
  \centering
  \includegraphics[width=\textwidth]{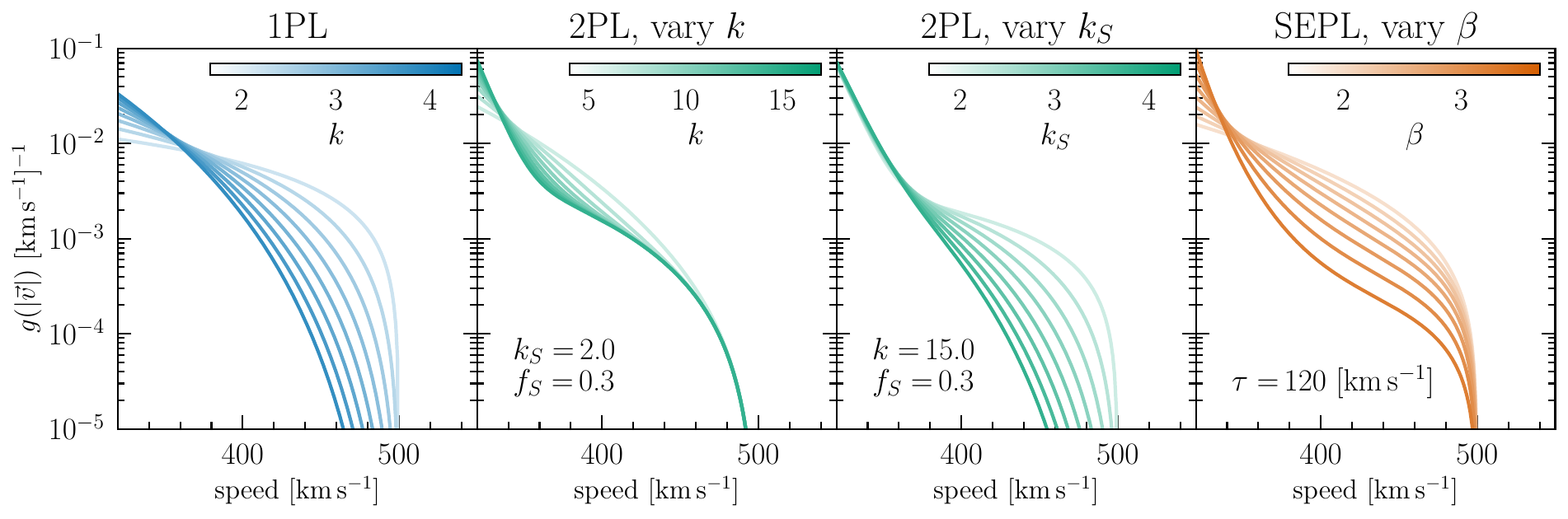}
  \caption{Shape comparison of the single power law model (1PL), two-component power law model (2PL) and the stretched exponential power law model (SEPL). All models shown have $v_{\rm{esc}} = 500\,\rm{km}\,\rm{s^{-1}}$. In the 1PL model, the power law index $k$ describes both the steepness of the distribution away from $\vesc$ and the shape of the cutoff near $\vesc$, leading to highly correlated fits for $\vesc$ and $k$.  The 2PL model mitigates this with the introduction of an additional power law-component with index $k_S$. In the SEPL model, the parameter $\beta$ controls  the exponential rise of the distribution away from $\vesc$ while the behavior near $\vesc$ is a power law with index $\gamma=1$.
  }
  \label{fig:modelcomp}
\end{figure*}

In these works, and as described in Sec. \ref{sec:methods}, it has also been necessary to set some lower bound on the speed $\vmin$ in order to model only the tail of the distribution. $\vmin$ is typically chosen close to $300\,\rm{km\,s^{-1}}$ (but lower in the pre-\Gaia~era due to limited statistics) to balance the statistical constraining power of the sample and the fact that Eq. \ref{eq:singlePL} is expected to model the tail of the distribution only. The escape velocities obtained by these analyses are typically of the order $500\,\rm{km\,s^{-1}}$, and it is not clear that Eq. \ref{eq:singlePL} should hold over such a large range of speeds, due to the presence of kinematic substructure, disk contamination, and the fact that this expression is obtained perturbatively in a neighborhood of $\vesc$. Some of these issues were  addressed in \cite{Koppelman2021} and \cite{Necib2022} by implementing aggressive cuts on disk stars, and furthermore in \cite{Necib2022} by repeating the modeling at different $\vmin$, where it was demonstrated that the 1PL modeling was not stable to the choice of $\vmin$. 

Since there exist different choices for priors on $k$ and for $\vmin$ which affect the resulting $\vesc$, the 1PL model may be inappropriate for robust estimation of escape velocities. When strict priors on $k$ are not employed, the resulting escape velocities are highly uncertain due to the strong parameter correlations. As a result, alternative models have been investigated; this is the subject of the following Section.

\subsection{Multiple power law components}
\label{sec:mpl}

The Milky Way contains kinematic substructure such as the GSE \citep{2018Natur.563...85H,2018MNRAS.478..611B}, and this substructure comprises a significant portion of the speed distribution \citep{Necib:2019a}. It was demonstrated in \cite{grand:2019,Necib2022methods} that the presence of kinematic substructures such as the GSE may significantly bias the determination of the escape velocity, in particular if using the single power law model (Eq. \ref{eq:singlePL}) with a low $\vmin$ such as $300\,\rm{km\,s^{-1}}$. 

Motivated by these considerations, \cite{Necib2022methods} proposed to model the local speed distribution as a sum of multiple bound kinematic components, each described by Eq. \ref{eq:singlePL} with independent slope but common $\vesc$, allowing for more robust estimates of the escape velocity. Via Eq. \ref{eq:multiple_components}, the likelihood for a single star $\alpha$ assuming a sum of two power law components with common $\vesc$ takes the form
\begin{equation}
    \begin{aligned}
    \mathcal{L}_\alpha^{2}=&\,(1-f_\mathrm{out})\, \big[(1-f_S)\, p_\alpha\left(v_\alpha \mid v_\mathrm{esc}, k\right) \\
    &+ f_S\, p_\alpha\left(v_\alpha \mid v_\mathrm{esc}, k_S\right)\big] +f_\mathrm{out} \,p_\alpha^{\mathrm{out}}\left(v_\alpha \mid \sigma_{\mathrm{out}}\right)
    \end{aligned}
\end{equation}
where each $p_\alpha$ implicitly uses the simple power law model Eq. \ref{eq:singlePL}. $k_{S}$ and $f_S$ are the power law slope and fractional contribution of the substructure component, respectively. Hereafter we will often refer to the single power law and 2-component power law models as ``1PL" and ``2PL", respectively. 

Applying this approach to the \Gaia eDR3 dataset, \cite{Necib2022} found that at low $v_{\rm{min}}$ ($\sim 320\,\rm{km\,s^{-1}}$) a two-component model was largely favored over a single bound component, whereas at $v_{\rm{min}}\gtrsim 360\,\rm{km\,s^{-1}}$ the single component model was slightly preferred, but the different models were not meaningfully distinguishable via the AIC. The three-component model was also found not to be distinguishable from the two-component, and as a result in this article we consider at most 2 bound components. 

\subsection{Limitations of the 2PL model}
\label{sec:limitations_2pl}

While \cite{Necib2022} found that the 2PL model provided a better fit to the \Gaia eDR3 data, it remained the case that high values of $k$ were preferred for one of the bound components, and even with a high prior limit of $k = 20$ the marginal distribution for $k$ remained unconverged. The 3-component model also exhibited the same convergence issues, suggesting that the source of the problem is a systematic effect of the modeling close to $\vmin$, where there is a steeply rising distribution of stars. 

In Fig. \ref{fig:modelcomp}, we illustrate the 1PL and 2PL models for fixed $\vesc$. In particular, we show how the various slope parameters affect the resulting speed distributions. Here one can see that the 1PL model would be heavily constrained by a steep distribution close to $\vmin$. This would favor high $k$ and thus strongly inform the $\vesc$ estimate. The 2PL model takes a step towards describing these separate parts of the distribution near $\vmin$ and $\vesc$ by allowing for different power law behaviors. However, its flexibility saturates at about $ k \simeq 15$, indicating that continuing to increase the upper prior limit further will not be sufficient. 

Assuming the problem lies only in disk contamination or the inability to model substructure, one approach would be to increase $\vmin$ until this steeply rising feature disappears or until the distribution is indistinguishable from the 1PL fit, as made precise by the AIC (Eq. \ref{eq:AIC}).
However, if we wish to apply this analysis across many Galactocentric radii, this would require the tuning of $\vmin$ in each radial bin based on some arbitrary convergence or stability criterion. One could replace the $\vmin$ cut by only using the fastest $N$ stars in each radial bin as in \cite{Koppelman2021}, but the choice of an appropriate $N$ is again challenging. As discussed earlier, the number of stars in each radial bin is affected by the selection function and quality cuts such as parallax error, with far fewer stars in the $4-5\,\rm{kpc}$ bin compared to the $7-8\,\rm{kpc}$ bin, for example. Since the statistics vary greatly across Galactocentric radii, $N$ would again require a tuning in each bin.

Even with a high $\vmin$, both the 1PL and 2PL models can estimate potentially unreliable escape velocities past the data region (i.e. much faster than the fastest star observed) depending on the data features. For example, it is observed in \cite{Koppelman2021} that beyond the solar position, escape velocity estimates based on the 1PL model curiously {\emph{rise}} with Galactocentric radius, contrary to the expected shape in realistic potentials. The speed distributions are reported to become steeper with Galactocentric radius, and since $\vesc$ is positively correlated with the slope $k$, this leads to higher $\vesc$. This feature suggests some radially-dependent effect in the shape of the distribution which is not well-modeled by the single power law and is not accounted for by the statistical uncertainties of parameter estimation. 

Given these limitations, it is thus desirable to develop a model that can both closely track the cutoff of a continuous kinematic distribution (i.e., the fastest non-outlier star, accounting for measurement errors) and account for the steeply rising distribution at lower speeds, without tuning $\vmin$. 

\subsection{Stretched exponential power law (SEPL)}
\label{sec:sepl}

We now introduce a new functional form, motivated by the challenges of fitting the data to a single or two-component power law model. In both cases, the data tends to be fit by increasingly high power law slopes $k$ unless a prior range or strict $\vmin$ are set.  These steep power laws can lead to unreliable extrapolation past the data region and large $\vesc$ estimates. The fundamental reason is that the steepness of the distribution is correlated with the shape of the cutoff near $\vesc$ in power law distributions.  This indicates the need for a model where the steep rise in the distribution away from $\vesc$ is not tied to the shape of the cutoff close to $\vesc$.

We use these facts as motivation to introduce a model that \textit{grows exponentially away} from $v_{\rm{esc}}$, but maintains the characteristics of a power law close to $v_{\rm{esc}}$. This is similar to the motivation for the 2PL model, but allows for a steeper rise at lower speeds. A general distribution satisfying these requirements is 
\begin{equation}
    g(v \mid \mathbf{\theta}) \propto (v_\mathrm{esc} - v)^\gamma \exp\left[\left(\frac{v_\mathrm{esc} - v}{\tau}\right)^\beta\right].
    \label{eq:generalSEPL}
\end{equation}
Close to $\vmin$, this distribution rises as a stretched exponential with index $\beta$ and scale $\tau$. Very close to $\vesc$, this distribution gives a power law with index $\gamma$, with lower $\gamma$ giving a more sharp cutoff at $\vesc$. Since we wish to avoid extrapolation of $\vesc$ beyond the data region, we find that choosing a low $\gamma = 1$ gives results for $\vesc$ that most closely track the fastest non-outlier star in the data region. We refer to this model with $\gamma=1$ as a ``stretched exponential power law"\footnote{The stretched exponential usually is written with a negative sign in the exponent, but we wish to model an exponential rise toward $\vmin$ away from $\vesc$, and so we make use of this name in a slightly non-standard way.} (SEPL), with the form
\begin{equation}\label{eq:SEPL}
    g(v \mid \mathbf{\theta}) \propto (v_\mathrm{esc} - v) \exp\left[\left(\frac{v_\mathrm{esc} - v}{\tau}\right)^\beta\right],
\end{equation}
where $v_\mathrm{esc}, \beta,$ and the scale $\tau$ are treated as free parameters. Here $\beta$ and $\tau$ have limited impact on the precise shape of the cutoff at speeds very close to $\vesc$, but allow for greater flexibility in describing the data at speeds where statistics are greatest. The choice of $\gamma=1$ in Eq. \ref{eq:SEPL} is further validated in Sec. \ref{sec:results} by repeating the analysis for $\gamma \in \{1,2,3\}$ and additionally with $\gamma$ as a free parameter with a wide uniform prior.

We demonstrate the shape of the SEPL model and compare it to the power law models in Fig.~\ref{fig:modelcomp}. Here we show a fixed scale $\tau$, although when fitting we leave this scale as a free parameter. The SEPL model mimics a valuable feature of the 2-component model, which is to describe the tail as the transition between two distinct behaviors (here exponential to power law), but more readily accounts for the steep rise, and ideally further reduces the correlation of any improperly converged parameters with $\vesc$. 

\subsection{Priors}
\label{sec:priors}

Since the choice of priors in the power law models is a subject of significant uncertainty, we adopt wide uniform priors on \textit{all parameters} of the bound component models, and in line with \cite{Necib2022methods}, we choose uniform priors on the logarithm of the outlier distribution parameters. A summary is given in Table \ref{tab:priors}. The prior limits of the power law models are also chosen in line with \cite{Necib2022methods}, where we maintain the choice that $k$ should label the behavior close to $\vmin$ in all cases, but remain agnostic to the interpretation of each component.

The prior limits in the case of the SEPL model are chosen such that any combination yields a distribution that can be reliably normalized according to Eq.~\ref{eq:normalizeprob}; the exponential nature can cause numerical overflow issues in some parts of the parameter space, though this does not represent any unphysical nature of the modeling. 
As a result, note that the maximum of the uniform escape velocity prior is lower than that of the power law models, but this upper limit is still significantly higher than existing estimates of $\vesc$ and is never informative for the resulting posterior distributions in this paper. 

\begin{deluxetable}{lcl}
\tablecaption{Prior distributions for the power law and stretched exponential power law models. The notation $\mathcal U (a,b)$ refers to a uniform prior distribution from $a$ to $b$ and $\mathcal U_{\,\rm{log}}(c,d)$ refers to a uniform prior in the log of that parameter from $c$ to $d$. \label{tab:priors}}
\tablewidth{\columnwidth}
\tabletypesize{\small}
\tablehead{
    \colhead{Model} &
    \colhead{Parameter} &
    \colhead{Prior}
}
\startdata
1-component & $\vesc$ & $\mathcal U (0, 1000)$ \\
power law (1PL) & $k$ & $\mathcal U (0.1, 20.0)$ \\
\midrule
2-component & $\vesc$ & $\mathcal U (0, 1000)$ \\
power law (2PL) & $k$ & $\mathcal U (0.1, 20.0)$ \\
& $k_S$ & $\mathcal U (0.1, k)$ \\
& $f_S$ & $\mathcal U (0, 1)$ \\
\midrule
Stretched exponential & $\vesc$ & $\mathcal U (0, 700)$ \\
power law (SEPL) & $\beta$ & $\mathcal U (1, 3.5)$ \\
& $\tau$ & $\mathcal U (80, 1000)$ \\
\midrule
Outlier model & $f_{\rm{out}}$ & $\mathcal U_{\,\rm{log}}(10^{-6},1)$ \\
& $\sigma_{\rm{out}}$ & $\mathcal U_{\,\rm{log}}(600,3000)$ \\
\enddata
\end{deluxetable}

\section{Escape velocity results} \label{sec:results}

\begin{figure*}
    \centering
\includegraphics[width=\textwidth]{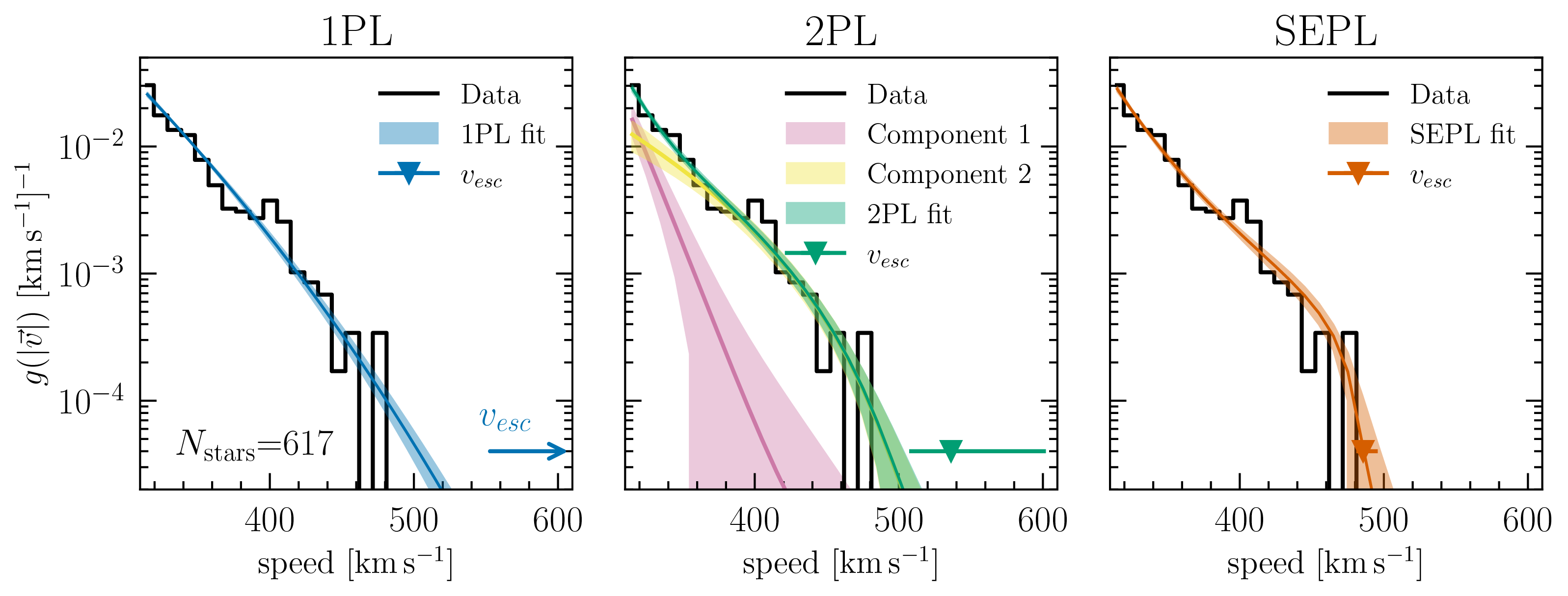}
    \caption{Comparison of fits to high-speed distribution of stars using the 1PL, 2PL and SEPL models, shown for a representative radial bin of $8-9\,\rm{kpc}$ and at a $\vmin$ of $310\,\rm{km\,s^{-1}}$. The fits at all Galactocentric radii can be found in the appendix. The arrow in the 1PL panel indicates that the escape velocity and its $68\%$ confidence interval are very large and not visible on these axes. The vertical position of the $\vesc$ markers does not encode any information.
    }
    \label{fig:modelcomp3panel}
\end{figure*}

We now apply the fitting procedure outlined in Sec. \ref{sec:methods} to the high-quality DR3 dataset of Sec. \ref{sec:data}, using the 1PL, 2PL and SEPL models of Sec. \ref{sec:modeling}. We perform the fit for each model in $1 \rm{kpc}$ radial bins from $4-11\,\rm{kpc}$. 

In Figure \ref{fig:modelcomp3panel}, we show the data and the best-fit speed distribution for each model, taking a representative radial bin of $8-9\,\rm{kpc}$ and $\vmin$ of $310\,\rm{km\,s^{-1}}$. The fits at all Galactocentric radii can be found in the Appendix, Fig.~\ref{fig:modelcompALL}. The corner plots corresponding to the fits at $8-9\,\rm{kpc}$ are also available in the Appendix (Figs. \ref{fig:corner1PL}, \ref{fig:corner2PL}, \ref{fig:cornerSEPL}).

In Figure \ref{fig:modelcomp3panel}, it can also be seen that the 1PL model results in a best-fit $\vesc$ which is well beyond the fastest star in the sample. Here the shape of the distribution at lower speeds is dictating the behavior of the high-speed tail, with an unreliable extrapolation of the distribution's shape to speeds where there is no data. The best fits for the 2PL model perform better, since the additional model parameters allow for more flexibility in modeling both the lower-speed and fastest stars. However, the 2PL model can also yield a $\vesc$ estimate far from the data region, relying again on strong assumptions about the shape of the distribution at high speeds. This effect is most pronounced for the power law models at large Galactocentric radius, as seen in Fig.~\ref{fig:modelcompALL}. Finally, the SEPL fits exhibit sharper cutoffs near $\vesc$, while simultaneously describing the steep rise of the distribution close to $\vmin$. The SEPL escape velocity estimate thus relies less on extrapolation of the speed distribution, and is very close to the observed cutoff in the distribution. The advantages of using the SEPL are discussed in more detail in Sec. \ref{sec:sepl_advantages}.

\begin{figure*}
    \centering
    \includegraphics[width=\textwidth]{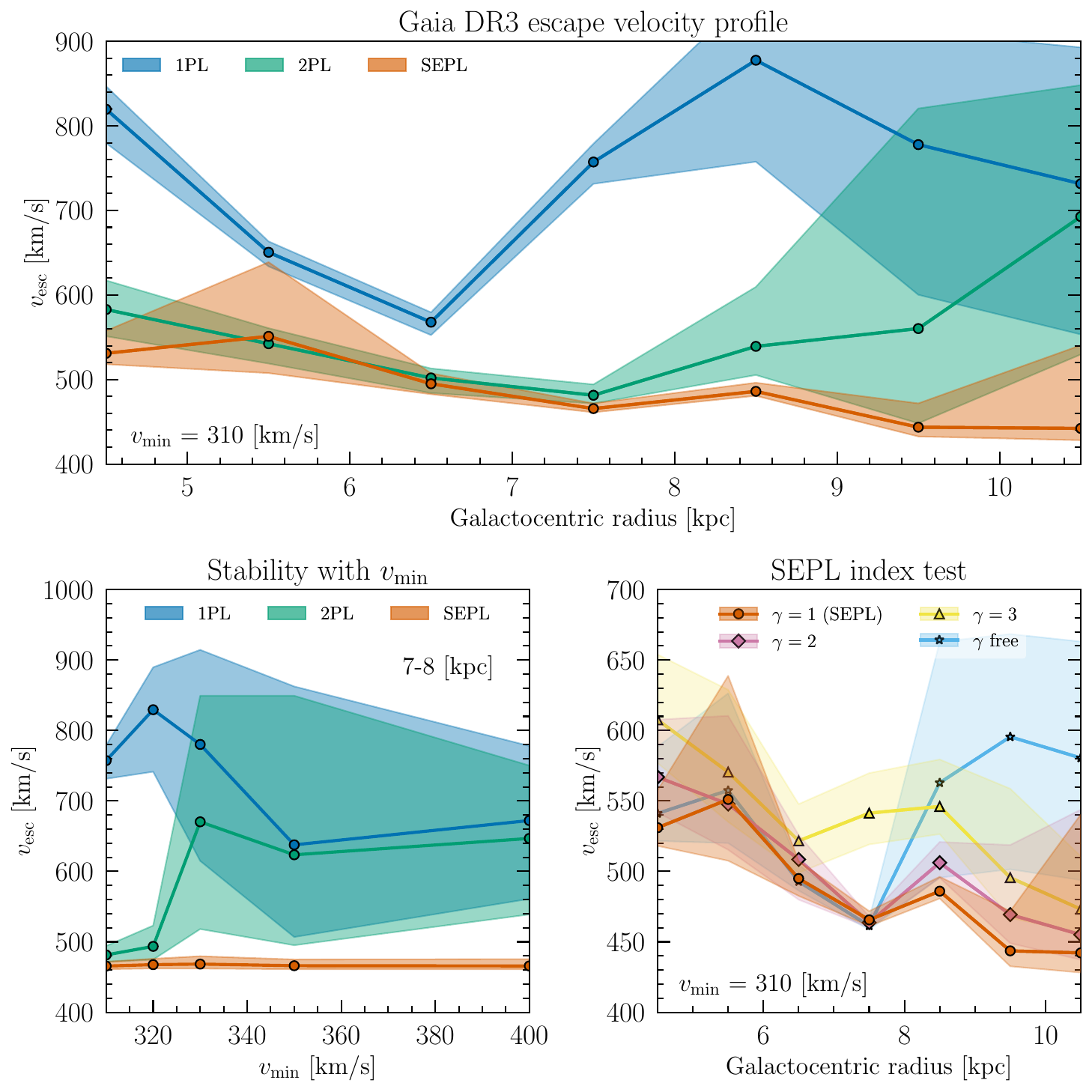}
    \caption{(Top) The escape velocity profile of the high-quality \Gaia~DR3 sample as obtained via the single power law (1PL), 2-component power law (2PL) and stretched exponential power law (SEPL) models, each at $\vmin=310\,\rm{km\,s^{-1}}$. (Bottom left) The stability of each model's escape velocity estimate with increasing $\vmin$, shown here for the representative radial bin of $7-8\,\rm{kpc}$. (Bottom right) Repeating the analysis with the SEPL model using different power law exponents $\gamma$ at high-speed extreme, including the case in which $\gamma$ is treated as a fit parameter. All bands correspond to $68\%$ confidence intervals.}
    \label{fig:3panelprofiles}
\end{figure*}

These features are also apparent when comparing the escape velocity profile of each model, as in the top panel of Fig.~\ref{fig:3panelprofiles}. The 1PL model confidently overestimates the escape velocities, and increases unphysically with Galactocentric radius from $6-9\,\rm{kpc}$. The 2PL model estimates lower escape velocities and relies less on an unreliable extrapolation beyond the data region, at least at small Galactocentric radius. However, at and beyond the Solar position, the 2PL fit produces similar results to the 1PL model and the escape velocity estimates become poorly constrained. The SEPL fitting typically results in the lowest estimates for the escape velocity, as the distribution can exhibit a sharper cutoff at the edge of the data region. Unlike the profiles of the power law models, the SEPL profile is consistent with monotonically decreasing with radius, which is a necessary feature of a physical escape velocity profile. The SEPL escape velocities are most uncertain when it is challenging to distinguish between bound and outlier distributions as in the $5-6\,\rm{kpc}$ bin, and when statistics degrade as at high Galactocentric radius.  

\subsection{Advantages of the SEPL model}
\label{sec:sepl_advantages}

Using the AIC (Eq. \ref{eq:AIC}) to compare the different models reveals that the SEPL and 2PL are not statistically distinguishable ($|\Delta \rm{AIC}|\lesssim 5$) and that the 1PL is strongly disfavored ($|\Delta \rm{AIC}| \simeq 10-100$). Despite the fact that the SEPL and 2PL have similar measures of goodness-of-fit, there are a number of reasons to favor the SEPL as our fiducial result. We discuss below the advantages of using the SEPL model, but note that in our final results for MW halo properties we will also consider $\vesc$ obtained with 2PL model to illustrate the model dependence.

As discussed above, the SEPL model gives results for $\vesc$ that rely less on extrapolation beyond the data region. This is preferable to the overestimation seen in the power law models (on the order of hundreds of $\rm{km\,s^{-1}}$ at high Galactocentric radius) as it is an approach informed more by the data than assumptions about the tail shape. In light of this discussion, one might ask whether it is appropriate to use the speed of the fastest star as an estimator for $\vesc$. The difficulty of this approach is that it does not account for outliers or statistical fluctuations near the tail, and degrades significantly in low statistics regimes such as at high Galactocentric radius. Using the SEPL model with a cutoff in the distribution accounts for these effects, and results in larger uncertainties on $\vesc$ in radial bins with limited statistics or outliers.

In addition, using the SEPL model mitigates one of the issues seen in power law models: the power law slope that governs the low-speed behavior, $k$, is often both unconverged and correlated with $\vesc$. In the Appendix, Fig.~\ref{fig:correlation} shows the degree to which the model parameters of the 1PL, 2PL and SEPL are correlated with $\vesc$ as a function of Galactocentric radius. In both the 1PL and 2PL models, $k$ exhibits a high correlation coefficient with $\vesc$, and it can be seen in the representative corner plots Fig. \ref{fig:corner1PL} and \ref{fig:corner2PL} that this parameter is unconverged even with a large prior range. This suggests that the marginal distribution for $\vesc$ is in fact not reliable in this case, as it is correlated with an unconverged parameter. We find the $\beta$ parameter in the SEPL model is sometimes similarly unconverged, in part because the upper prior limit is set to avoid numerical overflow. However, it is \textit{uncorrelated with the escape velocity} and thus does not pose the same problem as the power law models. There is one radial bin ($5-6\,\rm{kpc}$) in which there is meaningful correlation between $\beta$ and $\vesc$, but this is accompanied by a large uncertainty on the $\vesc$ estimate of the SEPL, and so does not significantly bias results of DM halo parameter fitting in Sec.~\ref{sec:DM}.

The lower left panel of Fig.~\ref{fig:3panelprofiles} shows the result of increasing $\vmin$ for the three models considered in this paper. For the power law models, the results for $\vesc$ become highly uncertain at high $\vmin$ as the statistics of the sample degrade, with both models estimating an escape velocity $\sim 150\,\rm{km\,s^{-1}}$ above the SEPL and low-$\vmin$ 2PL measurements. 
We find that the SEPL model gives results that are more stable to the choice of $\vmin$ than the power law models. This can be explained by the greater flexibility in the SEPL for describing the rise toward $\vmin$, which better describes the features observed in the data. Using the AIC to compare the models, the SEPL and 2PL results are indistinguishable at all $\vmin$ while all three models (1PL, 2PL, SEPL) become indistinguishable for high $\vmin$, consistent with expectations given the limited statistics. 

We also consider the more general form of the SEPL, Eq.~\ref{eq:generalSEPL}, with various choices of the power law index $\gamma$. The lower right panel of Fig.~\ref{fig:3panelprofiles} shows fit results with different fixed $\gamma=2,3$ as well as with $\gamma$ as a free parameter.  Larger $\gamma$ values give distributions with shallower cutoffs in the tail, which tend to result in higher $\vesc$ values. The SEPL with $\gamma$ a free parameter reproduces the behavior of the 2PL model close to $\vesc$, in particular rising and becoming more uncertain past the solar position. This feature could be due to the reduced number of stars above $\vmin$ in these bins, and the ``overshooting" issue of the power law models discussed in Section~\ref{sec:limitations_2pl}. We use $\gamma=1$ since it results in the lowest values of $\vesc$ in this comparison, consistent with our desire to avoid extrapolating beyond the data region. Further reducing $\gamma$ below 1 to give an even sharper cutoff would not substantially impact our results, since there are limited statistics to distinguish such low values of $\gamma$. For example, comparing the $\gamma=1$ and $\gamma=2$ cases we see these profiles are already largely consistent within $1\sigma$, but with the $\gamma=2$ case systematically above $\gamma=1$.

Lastly, we test the consistency of the escape velocity profile with an alternative radial binning (but the same bin width), offset by $0.5\,\rm{kpc}$ from the fiducial bins. This is shown in the Appendix Fig. \ref{fig:offsetbins} for both 2PL and SEPL models. The offset escape velocity profiles exhibit the same features as the fiducial profiles, confirming that bin membership uncertainty has a small effect on the resulting escape velocity results. 

\begin{figure*}[t]
    \centering
    \includegraphics[width=\textwidth]{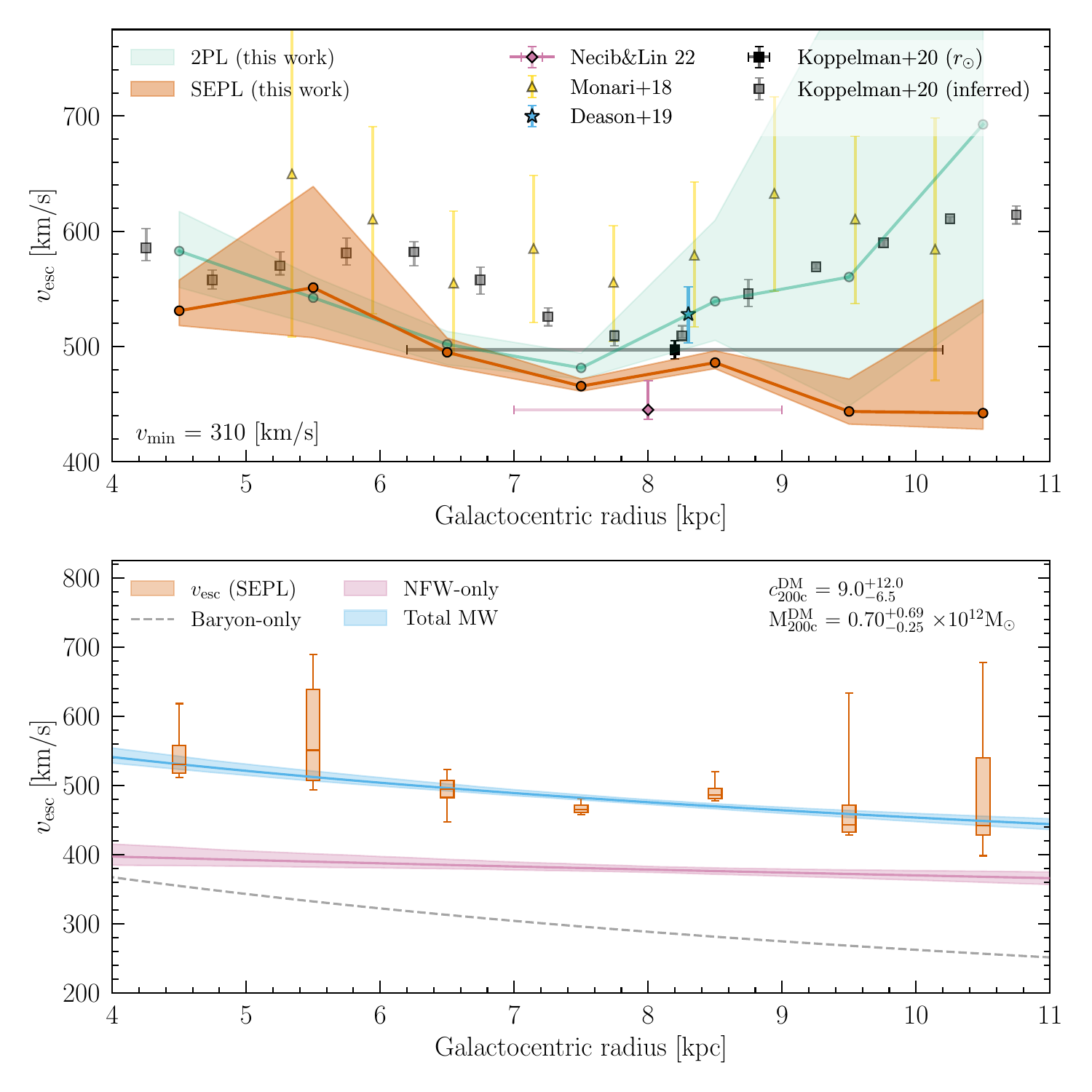}
    \caption{
    (Top) stretched exponential power law (SEPL) and 2-component power law (2PL) escape velocity profiles along with comparable results from the literature. Bands and error bars correspond to $68\%$ confidence intervals. Horizontal error bars for isolated results represent the Galactocentric radius range of sources used for that result. The escape velocities shown from the literature do not include any corrections motivated by simulations. (Bottom) Best fit of the NFW dark matter halo to the escape velocity profile, using SEPL escape velocity profile and assuming a known baryonic content of the Milky Way. The NFW halo parameters $\rm{M}_{200c}^{\rm{DM}}$ and $c_{200c}^{\rm{DM}}$ represent the virial mass and concentration of the dark matter halo. The SEPL profile is shown at $68\%$ (boxes) and $95\%$ (whiskers) confidence intervals. The baryon-only and NFW-only profiles show the escape velocity if the MW consisted of only those components, respectively. }
    \label{fig:litcomp_and_DM_profiles}
\end{figure*}

\subsection{Comparison with previous work}

For our final results, we consider the 2PL and SEPL models with the fiducial radial bins, and a default power law index of $\gamma =1$ for SEPL. We take $\vmin=310\,\rm{km\,s^{-1}}$ as the default because this maintains a large sample size while still remaining above the typically chosen $300\,\rm{km\,s^{-1}}$ to avoid disk contamination. Furthermore, as discussed above, the power law models perform less well at high $\vmin$ given the lower statistics. 

In Figure \ref{fig:litcomp_and_DM_profiles}, we present the comparison of the 2PL and SEPL escape velocity profiles with individual measurements and profiles from the literature. We find an escape velocity of $486^{+10}_{-5}\,\rm{km\,s^{-1}}$ for the SEPL model and $539^{+69}_{-34}\,\rm{km\,s^{-1}}$ for the 2PL model in the $8-9\,\rm{kpc}$ bin, which is largely consistent with previous work obtaining $\vesc$ near the solar radius. For the \cite{Necib2022} result, the escape velocity is consistent within $1\sigma$ for the SEPL measurement and $\sim 1.5\sigma$ for the 2PL measurement, but the escape velocity found in this work is larger by $\sim20-30\,\rm{km\,s^{-1}}$. This is likely higher due to the larger statistics of \Gaia~DR3 as compared to eDR3, allowing for stricter quality cuts and thus a more reliable discrimination between outlier and bound distributions, in addition to speed distributions which are filled closer to the escape velocity.

Compared to the measurements of \cite{deason:2019} and \cite{Koppelman2021} near the solar radius, our $\vesc$ estimates are consistent. Note that the values cannot be compared directly, since the analysis of \cite{deason:2019} utilizes an ansatz about the shape of the escape velocity profile, and the results of \cite{Koppelman2021} are obtained using a 4 kpc radial bin (horizontal bar), compared to our results in 1 kpc bins. The $\vesc$ parameter is known to be correlated with the fastest star in the distribution being fit, and if using a wide radial bin, one will likely preferentially fit the high-speed stars which are expected to come from the smallest radii. Examining the horizontal error bars of Fig. \ref{fig:litcomp_and_DM_profiles}, it is plausible that the \cite{Koppelman2021} measurement is more appropriately compared with the 2PL or SEPL measurements at the lower radii of $\sim6-7$ kpc.

Turning to the escape velocity profile, the key feature of the SEPL escape velocity profile is that it trends toward lower $\vesc$ at large Galactocentric radii, consistent with expectation. This is in contrast to fits with power law models, which exhibit a significant rise at and beyond the solar position. Our results with the 2PL model are similar to the profile of \cite{Koppelman2021}, which exhibits approximately the same rise in $\vesc$ at large Galactocentric radii. Note that in \cite{Koppelman2021}, in order to obtain what we label as the ``inferred profile" (Fig. \ref{fig:litcomp_and_DM_profiles}), first a fit is performed using stars within $2\,\rm{kpc}$ of the solar position. Then using the posterior for $k$ in this local fit as a {\emph{prior}} on $k$, escape velocity fits are obtained in $0.5\,\rm{kpc}$ bins from $4-11\,\rm{kpc}$, giving an escape velocity profile assuming the power law slope is consistent across Galactocentric radii. We also show the profile of \cite{Monari2018}, which was obtained using \Gaia~DR2 with independent fits to $\vesc$ and $k$ in radial bins, using a bootstrap method to account for uncertainties in bin membership and speed. It does not exhibit a significant trend due to the large uncertainties resulting from the relatively small statistics of \Gaia~DR2, but also gives high $\vesc$ estimates at large Galactocentric radii. These features suggest that power law modeling does not reliably describe the high-speed tail, at least at larger radii.

Note that \cite{williams:2017} also fit an escape velocity profile using data from the Sloan Digital Sky Survey \citep{SDSS:2012}, but the modelling was performed in a distinct manner from this work and the other papers discussed in this section. In particular, \cite{williams:2017} makes an ansatz for the radial dependence of the escape velocity profile, fits the ansatz to data over a wide range of radial values simultaneously, and then uses the result to obtain the escape velocity at the solar position. The paper infers a local escape velocity of $\vesc(r_\odot) = 521^{+46}_{-30}\,{\rm{km\,s^{-1}}}$, consistent with previous estimates at that position \citep{Monari2018, deason:2019, Koppelman2021} and the  SEPL and 2PL inference of this work at the $\sim 1\sigma$ level, and consistent with the local estimates of \cite{Necib2022} at the $\sim 2\sigma$ level. Additionally, in \cite{Prudil:2022} the 1PL modelling approach was applied to speed data from RR Lyrae stars from $4-12\,{\rm{kpc}}$ to obtain a measurement of $\vesc(r_\odot) = 512^{+94}_{-37}\,{\rm{km\,s^{-1}}}$, which is consistent with all measurements discussed here, likely in part due to the large range in Galactocentric radii used for the inference.

\section{Milky Way Halo Constraints}\label{sec:DM}

\begin{deluxetable*}{lcllr}
\tablecaption{Chosen values for the modeling of different Milky Way components, for the purpose of obtaining a mass profile from an escape velocity profile. The notation $\mathcal{U} (a,b)$ refers to a uniform prior distribution from $a$ to $b$. \label{tab:MW_components}}
\tablewidth{0.75\textwidth}
\tabletypesize{\small}
\tablehead{
    \colhead{Component} &
    \colhead{Model} &
    \colhead{Parameter} &
    \colhead{Symbol} &
    \colhead{Value}
}
\startdata
Bulge & Plummer profile & Mass & $\rm{M}_{\text{bulge}}$ & $1.067\times 10^{10}\,\rm{M}_\odot$ \\
& \citep{plummer:1911} & Scale radius & $b$ & $0.3\,\rm{kpc}$ \\
\midrule
Thin disk & Miyamoto-Nagai profile & Mass & $\rm{M}_{\text{thin disk}}$ & $3.944\times 10^{10}\,\rm{M}_\odot$ \\
& \citep{miyamoto:1975} & Scale radius & $r_{\text{thin disk}}$ & $5.3\,\rm{kpc}$ \\
& & Scale height & $z_{\text{thin disk}}$ & $0.25\,\rm{kpc}$ \\
\midrule
Thick disk & Miyamoto-Nagai profile & Mass & $\rm{M}_{\text{thick disk}}$ & $3.944\times 10^{10}\,\rm{M}_\odot$ \\
& \citep{miyamoto:1975} & Scale radius & $r_{\text{thick disk}}$ & $2.6\,\rm{kpc}$ \\
& & Scale height & $z_{\text{thick disk}}$ & $0.8\,\rm{kpc}$ \\
\midrule
DM halo & NFW profile & Mass & $\rm{M}_{200c}^{\rm{DM}}$ & $\mathcal U (10^{10}, 10^{13})\rm{M}_\odot$ \\
& \citep{navarro:1996} & Concentration & $c_{200c}^{\rm{DM}}$ & $\mathcal U (0.1, 50.0)$ \\
\enddata
\end{deluxetable*}

\subsection{From escape velocity to dark matter}

The escape velocity profile of an isolated, finite, and spherical mass distribution is related to its potential $\Phi(\vert \vec{r} \vert)$ via $\vesc (\vert \vec{r} \vert) = \sqrt{2\Phi (\vert \vec{r} \vert)}$. In practice, however, the Milky Way is neither isolated nor spherical. It is therefore necessary to choose a distance at which we consider a star to be unbound, the choice of which is somewhat arbitrary. To remain consistent with the literature \citep{deason:2019,Necib2022}, we choose $2R_{200\rm{c}}$ as this limiting radius, where $R_{200\rm{c}}$ is the radius within which the galaxy’s mean density is 200 times the critical density of the universe
\begin{equation}
    \rho_{\rm{c}} = \frac{3 H^2}{8\pi G},
\end{equation}
where $H$ is the Hubble constant, and $G$ is the Newton's constant of gravity.
In this work, we adopt a Hubble constant of $H = 70\,\rm{km\,s^{-1}\,Mpc^{-1}}$ (although slightly different from recent measurements~\citep{ade:2016}, this value is chosen for consistency with prior analyses). With this assumption, the escape velocity is related to the gravitational potential via 
\begin{equation}
    \vesc (r) = \sqrt{2\vert \Phi (r) - \Phi(2R_{200\rm{c}})\vert}
\end{equation}
which we evaluate with cylindrical radius ($r$) in the plane of the disk for axisymmetric potentials. One can then obtain the relationship between the potential and density profile via Poisson's equation, establishing the link between $\vesc(r)$ and $\rho(r)$. 

To recover the density profile of the DM, it is necessary to model the density profiles of each of the Milky Way components, namely the bulge, thin disk, thick disk, and DM halo. In this work, we adopt model I of \cite{pouliasis:2017}, which we show in Table \ref{tab:MW_components}, to stay consistent with previous studies \citep{deason:2019, Necib2022}.
 
We consider the Navarro-Frenck-White (NFW) profile \citep{navarro:1996, navarro:1997} for the DM component, with density profile given by 
\begin{equation}
    \label{eq:NFW}
    \rho(r)=\frac{\rho_0}{\frac{r}{R_s}\left(1+\frac{r}{R_s}\right)^2},
\end{equation}
where $\rho_0$ is the normalization and $R_s$ is the scale radius. This profile can be formulated in terms of the equivalent pair of parameters $\rm{M}_{200\rm{c}}^{\rm{DM}}$ (DM mass within a sphere of radius $R_{200\rm{c}}$) and concentration $c_{200\rm{c}}^{\rm{DM}} = R_{200\rm{c}}/R_s$. The relation between these parameters is given by integrating the density profile to obtain
\begin{align}
\begin{split}
    \label{eq:NFW_conversion} {\rm{M}}_{200\rm{c}}^{\rm{DM}}&=\int_0^{R_{200{\rm{c}}}} 4 \pi r^2 \rho(r) d r\\
    &=4 \pi \rho_0 R_s^3\left[\ln (1+c)-\frac{c}{1+c}\right]
\end{split}
\end{align}
wherein $c$ is used as a shorthand for $c_{200\rm{c}}^{\rm{DM}}$. The priors on $\rm{M}_{200\rm{c}}^{\rm{DM}}$ and $c_{200\rm{c}}^{\rm{DM}}$ used in fitting are shown in Table \ref{tab:MW_components}.

\subsection{Milky Way Halo Parameters}
We now obtain the halo parameters for the Milky Way by simultaneously fitting the entire escape velocity profile for a given speed distribution model, such as the SEPL or 2PL. We use the full $\vesc$ posterior at each Galactocentric radius to evaluate the likelihood of a given set of DM halo parameters at that radius, and assume that each measurement in the profile is independent such that the total likelihood is the product of the individual evaluations of the posteriors. The theoretical escape velocity values at each radius for a given $\rm{M}_{200\rm{c}}^{\rm{DM}}$ and $c_{200\rm{c}}^{\rm{DM}}$ are obtained by modeling the galaxy as the combination of this NFW halo and the baryonic components shown in Table \ref{tab:MW_components}, performed using \texttt{galpy} \citep{2015ApJS..216...29B}. 

\begin{figure*}[t]
    \centering
    \includegraphics[width=\linewidth]{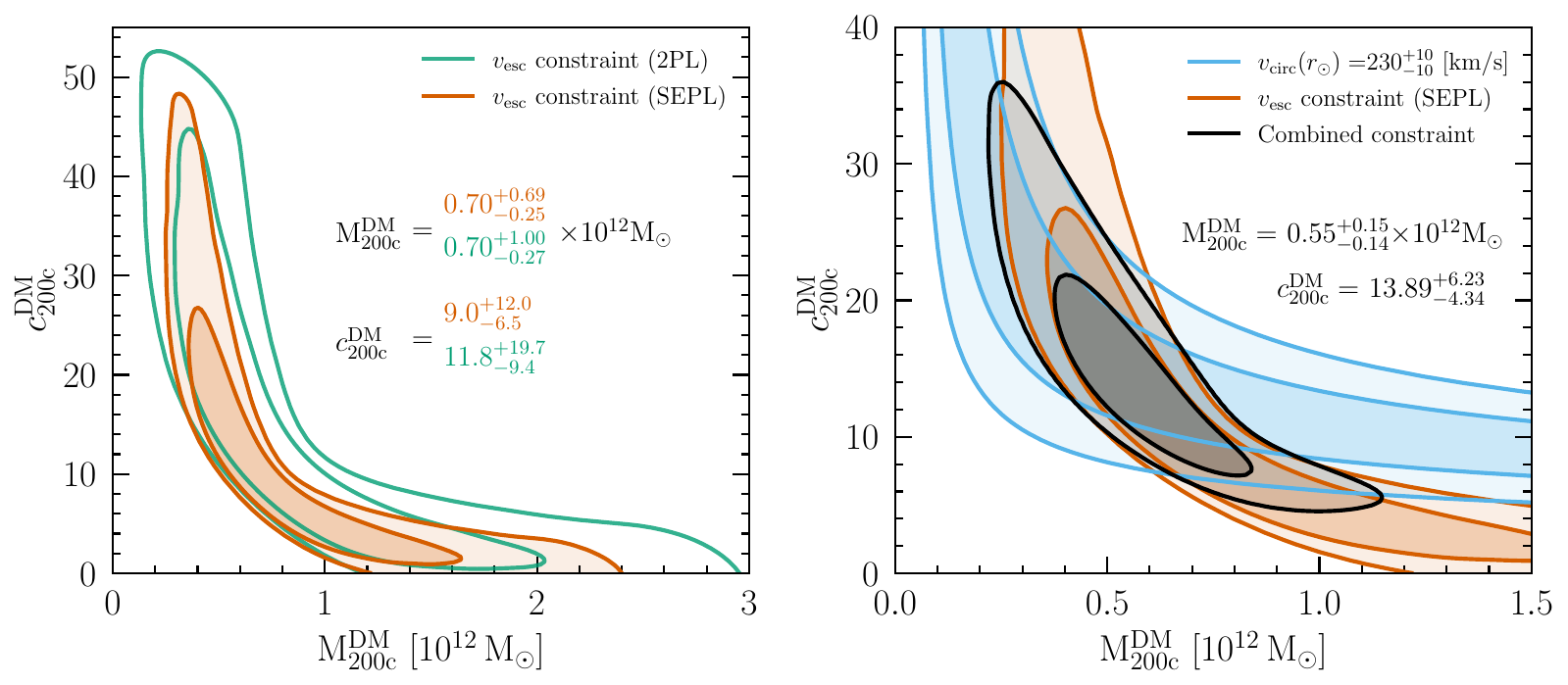}
    \caption{Left: Constraints on the parameter space of the NFW halo of the Milky Way, as obtained via the escape velocity measurements of both the two-component power law (2PL) and stretched exponential power law (SEPL) models. Right: Combined constraints on the Milky Way NFW halo parameters due to the SEPL escape velocity constraints and the circular velocity constraint at the solar position of \cite{eilers:2019}. Contours correspond to $68\%$ and $95\%$ confidence.}
    \label{fig:NFW_param_constraints}
\end{figure*}

We perform the fits using \texttt{emcee} \citep{2013PASP..125..306F} which is also an affine-invariant Markov Chain Monte Carlo sampler. The best fit to the SEPL escape velocity profile is shown in Fig. \ref{fig:litcomp_and_DM_profiles}, and the shaded band is the 68\% confidence interval. We also show the decomposition of the best fit profile into the baryon-only and NFW-only components. 
Constraints on the NFW parameter space due to the $\vesc$ measurements of both the SEPL and 2PL models are shown in Fig. \ref{fig:NFW_param_constraints}. It can be seen that both models yield consistent constraints, but those of the 2PL model are more uncertain and biased toward slightly higher masses, likely due to overestimation and the large uncertainty on $\vesc$ beyond the solar position for this model. 

To obtain stronger constraints on the NFW parameter space, previous works have combined escape velocity constraints with those of circular velocity measurements \citep{Piffl2014, Monari2018, deason:2019, Koppelman2021, Necib2022}. 
This is because the escape velocity probes the large-scale mass of the Milky Way, whereas circular velocity measurements are sensitive to the enclosed mass at that position. With the same baryonic model from Table \ref{tab:MW_components}, in line with \cite{Necib2022} we also include the circular velocity measurement at the solar position of \cite{eilers:2019},\footnote{More recent studies, such as \cite{Ou:2023}, have produced new measurements of the circular velocity based on \Gaia DR3. However, we adopt the older study of \cite{eilers:2019} for two reasons: First, it is consistent with \cite{Necib2022}, enabling a more direct comparison, and second, recent studies have shown a significant decline in the rotation curve at large Galactocentric radii that remains to be addressed. We try to avoid such a result affecting our measurements until a full understanding of the new circular velocity curves is established.} namely $v_{\rm{circ}}(r_\odot) = 230^{+10}_{-10}\,\rm{km\,s^{-1}}$. Our combined $\vesc$ and $v_{\rm{circ}}$ constraints on the NFW parameter space are shown in Fig. \ref{fig:NFW_param_constraints}. 
We obtain a final mass estimate for the DM halo of $\rm{M}_{200\rm{c}}^{\rm{DM}} = 0.55^{+0.15}_{-0.14}\times 10^{12} \, M_\odot$, which corresponds to a total Milky Way mass of $\rm{M}_{200\rm{c}} = 0.64^{+0.15}_{-0.14}\times 10^{12} \, M_\odot$ with this baryonic model.

\begin{figure}[t]
    \includegraphics[width=\linewidth]{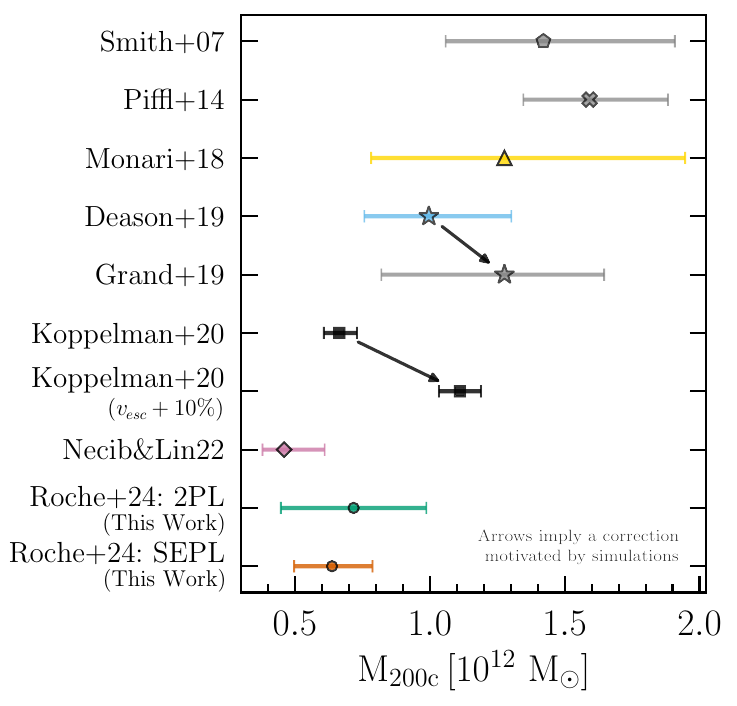}
    \caption{Milky Way total mass estimates obtained via escape velocity modeling along with  circular velocity constraints. We compare results from the literature with those of this work. All error bars correspond to $68\%$ confidence intervals, and arrows imply corrections to the inferred escape velocities motivated by simulations.}
    \label{fig:mw_mass_comp}
\end{figure}

A comparison of Milky Way mass estimates obtained by fitting both escape velocity and local circular velocity measurements is shown in Fig. \ref{fig:mw_mass_comp}. Here it can be seen that the mass estimates obtained via the SEPL modeling and 2PL modeling are consistent, though the 2PL results are less constrained. Furthermore, these results are consistent with those of \cite{Necib2022} which were obtained via a 2PL measurement of $\vesc$ within 2~kpc of the solar position using \Gaia eDR3 data. The results are also consistent with the uncorrected \cite{Koppelman2021} mass estimate; the correction is motivated by simulations and amounts to increasing all escape velocity estimates by $10\%$. Our results support the trend of  analyses in recent years based on escape velocity modeling, which tend to find a lighter Milky Way of mass $\sim 0.5-1\times 10^{12}\,M_\odot$. This is likely due to the ``overshooting" of the 1PL inflating $\vesc$ measurements in earlier work.

\section{Conclusions}
In this paper, we have measured an escape velocity profile of the Milky Way from $4-11\,\rm{kpc}$ using a sample of high-speed stars from \Gaia~DR3 with 6D kinematics and strict quality cuts. Following \cite{Necib2022methods}, we considered single and multi-component power laws to model the tail of the stellar speed distribution and extract escape velocities. Similar to \cite{Necib2022}, we found that a single power law fit does not provide a good fit to the tail and that it often significantly overestimates the escape velocity compared to the fastest star in the data. The multi-component power law model is motivated by the presence of kinematic substructure, and provides a much better fit. However, with the high quality data of \Gaia~DR3 and a large range of radii, we found even the two-component power law tends to systematically predict higher $\vesc$. This is particularly noticeable at positions beyond the solar radius, where the power law models give a rising escape velocity profile. Furthermore the $\vesc$ result is highly correlated with the slope of the distribution at low speeds, which is often an unconverged parameter in the fits.

Motivated by the above issues with the power law models, we introduced an empirical model which we call the ``stretched exponential power law" (SEPL). This model was motivated by the same features that multi-component power laws aimed to capture, such as a different behavior near $\vesc$ and near $\vmin$. However, in this case, there is more freedom for a steep rise in the distribution near $\vmin$, while the profile near $\vesc$ remains that of a power law. The escape velocity profile obtained using this new model is less susceptible to the issue of overestimating $\vesc$, and gives a falling profile past the solar position, consistent with expectation. This model is empirically motivated, and future work may improve this modeling as data quality and statistics improve or by studying its application to simulations.

Using the resulting escape velocity profile, we constrained the DM halo parameters of the Milky Way, in particular its virial mass, by assuming a model of the baryonic content consistent with the literature. By combining constraints with a complementary circular velocity measurement, we find a total virial mass for the Milky Way  of $M_{200\rm{c}} = 0.64^{+0.15}_{-0.14}\times 10^{12} \, \rm{M}_\odot$ using the SEPL model for speed distributions. We obtain consistent but less constrained estimates using the 2PL model. This Milky Way mass is found to be consistent with recent measurements in the literature using power law models, albeit on the lighter end, following the recent trend with such analyses. This trend may result from the improvement in data as well as more modeling techniques that account for substructure in the speed distribution. It has also been demonstrated in \cite{Kravtsov:2024} that a massive satellite galaxy such as the Large Magellanic Cloud can lead to the development of a high-velocity tail in the host, which would bias mass inferences made using stellar speeds. As a result, future modelling may require more sophisticated Milky Way system mass models, or may utilize comparisons to simulated Milky Way analogs including massive companions \citep{Buch:2024}. 

Apart from the escape velocity, the mass of the Milky Way has been inferred by numerous methods, each resulting in estimates distributed roughly between $0.5-2\times 10^{12}\,\rm{M}_\odot$ over the past decade \citep{wang:2020}. Masses inferred via escape velocities have constituted some of the larger estimates in this population in the past; however, as mentioned above, recent estimates trend below $10^{12}\,\rm{M}_\odot$. Interestingly, some measurements of masses inferred via other probes such as the rotation curve of the inner Milky Way \citep{Ou:2023, Francesco:2023, 2023arXiv230900048J, 2023ApJ...946...73Z} and comparing the Milky Way satellite population to simulations \citep{barber:2014, cautun:2014, patel:2018} have also trended downward in the past decade. This sits in contrast to some recent Galaxy mass estimates obtained using tracer stars \citep{Shen:2022} and satellite dwarfs \citep{fritz:2020, Slizewski:2022}, which suggest an intermediate-mass Milky Way ($\sim 1.4\times 10^{12}\,\rm{M_\odot}$). Other probes such as phase space distribution modeling \citep{li:2020, callingham:2019} and modeling the orbital dynamics of globular clusters \citep{sohn:2018,watkins:2019,Jianling:2022} exhibit no clear trend. See \cite{wang:2020} and \cite{Sawala:2023} for reviews of the Milky Way mass obtained by various methods. These results together paint a picture of a Milky Way whose mass is close to $10^{12}\,M_\odot$, although with some disagreement to within a factor $\lesssim 2$. As observations continue to rapidly improve, a unified understanding of these different facets of the Milky Way's dynamics will be essential to ultimately pinning down the characteristics and history of our Galaxy.

\section*{Acknowledgements}
This work has made use of data from the European Space Agency (ESA) mission
{\it Gaia} (\url{https://www.cosmos.esa.int/gaia}), processed by the {\it Gaia}
Data Processing and Analysis Consortium (DPAC,
\url{https://www.cosmos.esa.int/web/gaia/dpac/consortium}). Funding for the DPAC
has been provided by national institutions, in particular the institutions
participating in the {\it Gaia} Multilateral Agreement. TL acknowledges support from the US Department of Energy Office of Science under Award No. DE-SC0022104. The authors also thank Philip Harris for valuable discussions.

\software{
Python \citep{python}, 
numpy \citep{numpy:2020}, scipy \citep{scipy:2020}, 
astropy \citep{astropy:2013, astropy:2018}, 
jupyter \citep{jupyter}, 
corner.py \citep{corner}, emcee \citep{2013PASP..125..306F}, Galpy \citep{2015ApJS..216...29B}. The implementation of the affine-invariant Markov chain Monte Carlo algorithm of \cite{AIMCMC} in the \texttt{julia} programming language \citep{julia} can be found at \url{https://github.com/CianMRoche/MCJulia.jl} which is forked from \url{https://github.com/mktranstrum/MCJulia.jl}.}

\bibliography{bibliography}{}
\bibliographystyle{aasjournal}

\appendix
\renewcommand{\thefigure}{A\arabic{figure}}
\setcounter{figure}{0}
The Figures in this Appendix are organized as follows: Fig. \ref{fig:offsetbins} contains a test of an alternative radial binning to produce 2PL and SEPL escape velocity profiles. Fig. \ref{fig:modelcompALL} is a collection of the speed distributions at all Galactocentric radii considered here and the corresponding 1PL, 2PL and SEPL fits. Fig. \ref{fig:corner1PL}, \ref{fig:corner2PL} and \ref{fig:cornerSEPL} are the corner plots corresponding to the fits shown in Fig. \ref{fig:modelcomp3panel}. Fig. \ref{fig:correlation} shows the correlation between best-fit model parameters for the 1PL, 2PL and SEPL models and the escape velocity as a function of galactocentric radius. The correlation is calculated as the absolute value of the Pearson product-moment correlation coefficient using the \textsc{numpy} function \texttt{corrcoef} \citep{numpy:2020}.

\begin{figure*}[h!]
    \centering
    \includegraphics[width=\textwidth]{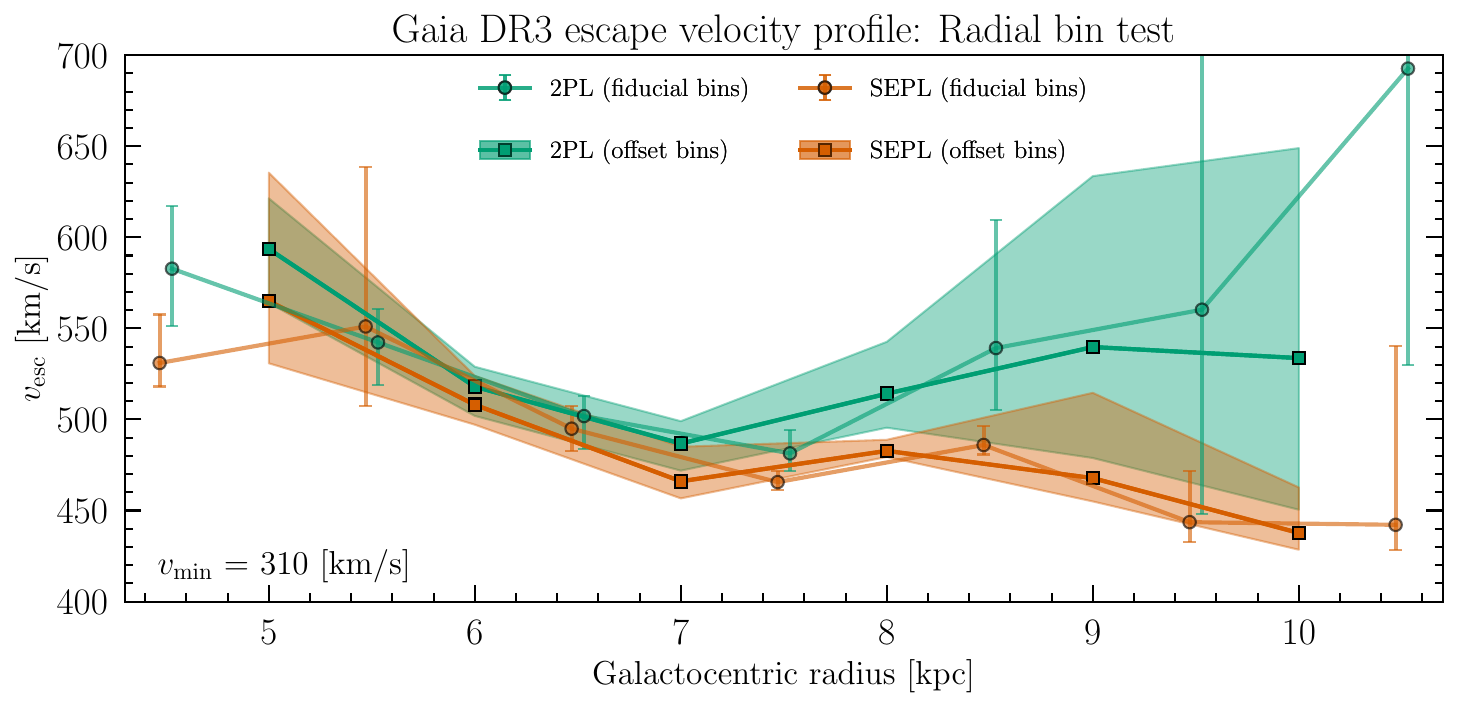}
    \caption{Stretched exponential power law (SEPL) and two-component power law (2PL) escape velocity fits to \Gaia~DR3 data from $4-11\,\rm{kpc}$ in the fiducial $1\,\rm{kpc}$-wide radial bins (circles, error bars) and in bins offset by $0.5\,\rm{kpc}$ (squares, bands), both with $\vmin=310\,\rm{km\,s^{-1}}$. Error bars and bands represent $68\%$ confidence intervals.}
    \label{fig:offsetbins}
\end{figure*}

\begin{figure*}
    \centering
    \includegraphics[height = 0.9\textheight]{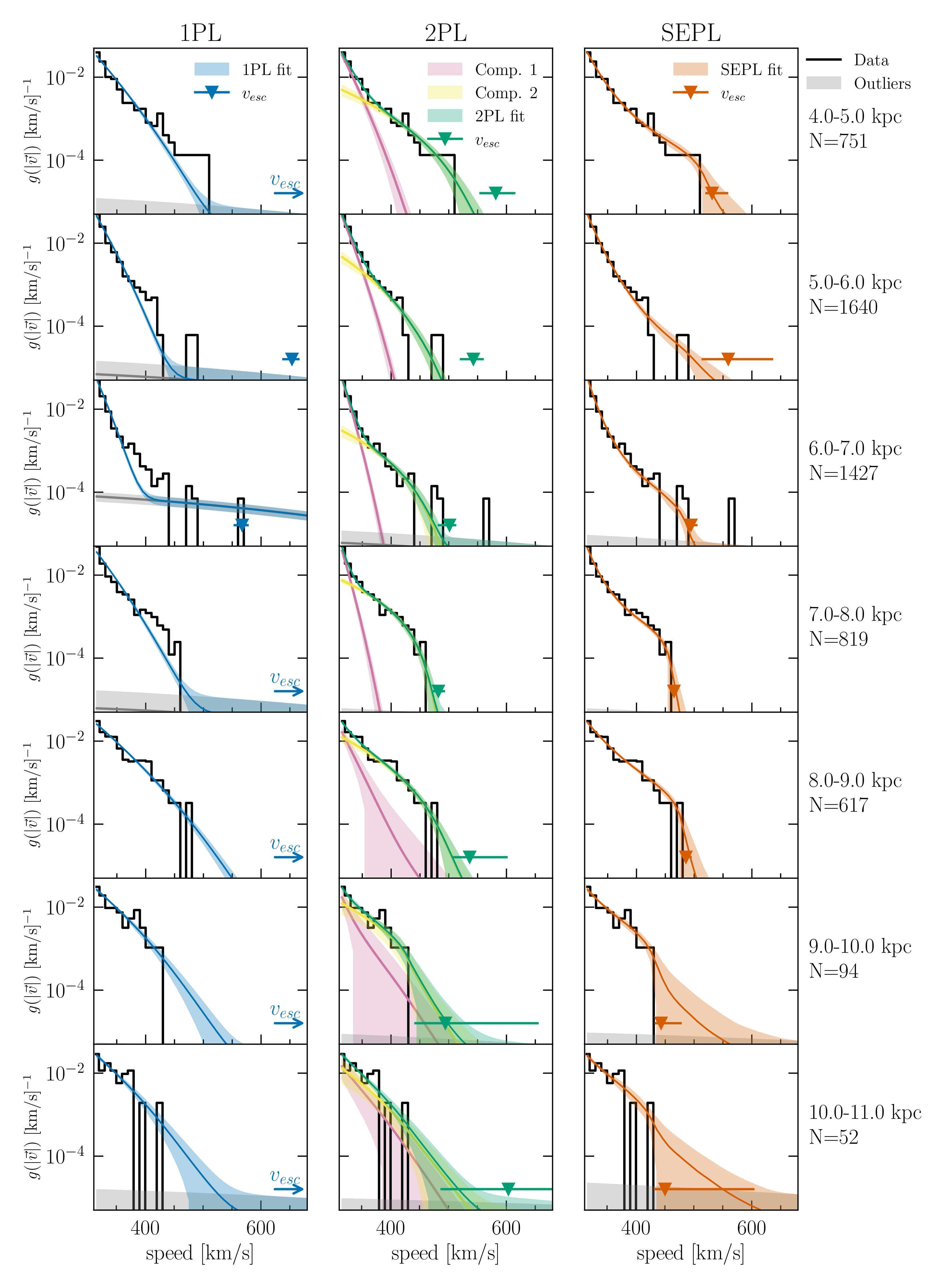}
    \caption{Escape velocity fits using the power law (1PL), two-component power law (2PL) and stretched exponential power law (SEPL) at $\vmin = 310\,\rm{km\,s^{-1}}$ at all Galactocentric radii considered in this work. All bands and error bars represent $68\%$ confidence intervals. Vertical position of $\vesc$ markers do not encode any information.} 
    \label{fig:modelcompALL}
\end{figure*}

\begin{figure*}
    \centering
    \includegraphics[width=0.48\textwidth]{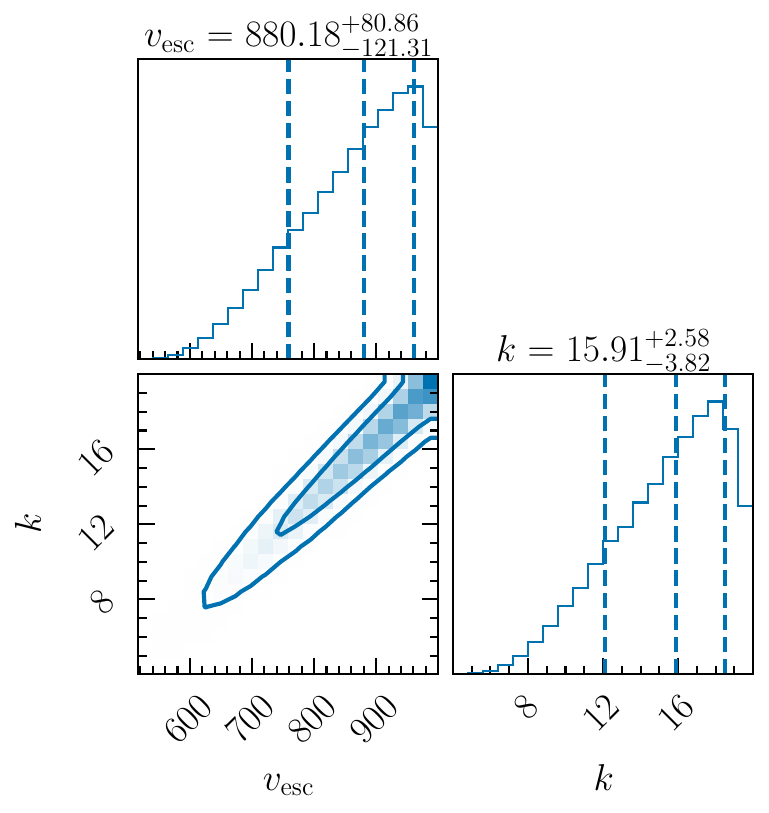}
    \includegraphics[width=0.48\textwidth]{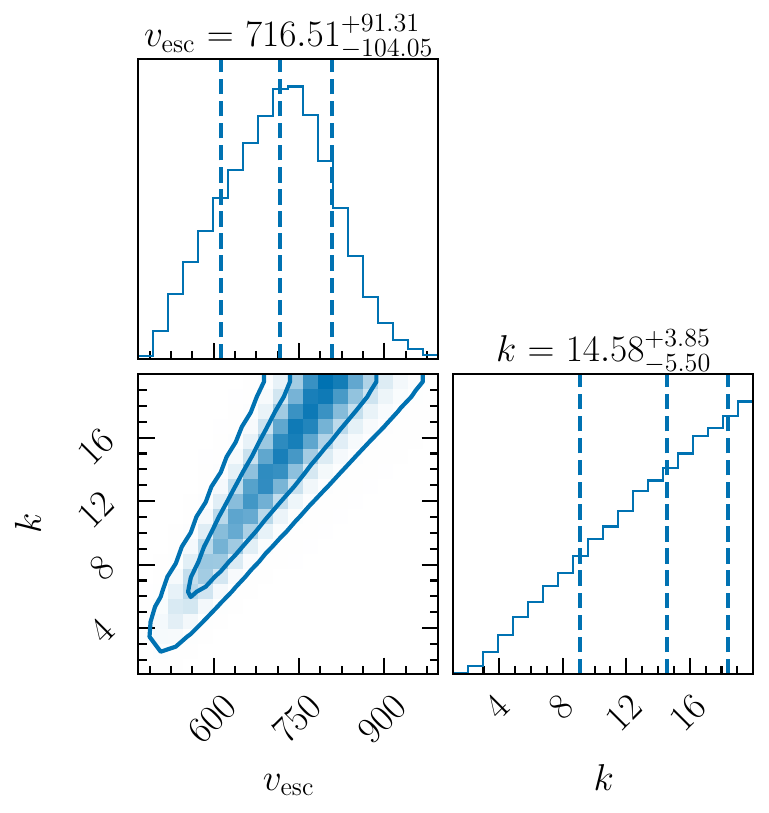}
    \caption{
    Single power law (1PL) fit to  \Gaia~DR3 data in the $8-9\,\rm{kpc}$ radial bin and with $\vmin=310\,\rm{km\,s^{-1}}$ (left) and $\vmin=400\,\rm{km\,s^{-1}}$ (right). 
   We do not show the outlier fraction $f_{\rm{out}}$ and distribution width $\sigma_{\rm{out}}$, as the outlier distribution is consistent with zero in this radial bin. }
    \label{fig:corner1PL}
\end{figure*}

\begin{figure*}
    \centering
    \includegraphics[width=\textwidth]{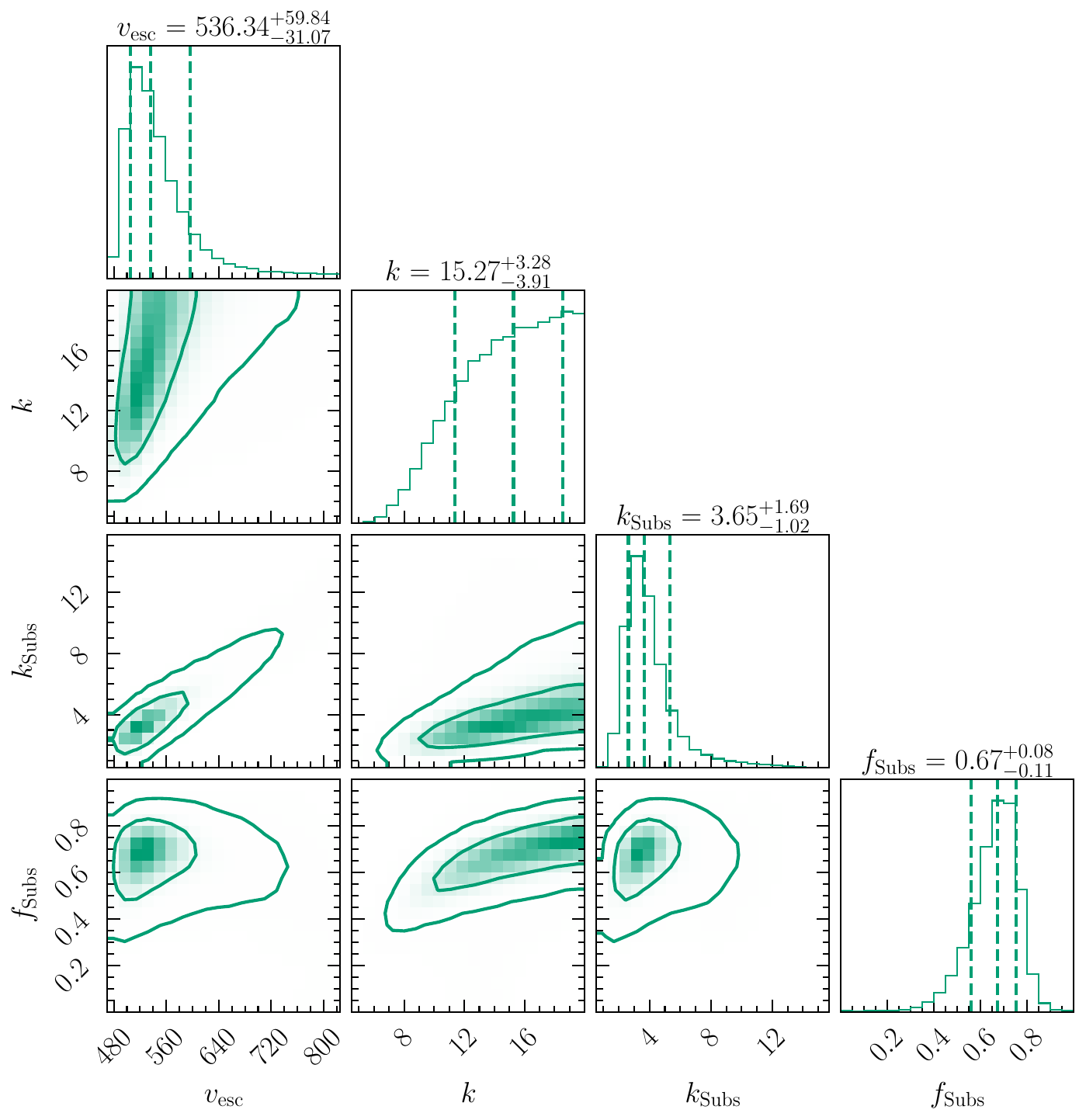}
    \caption{Two-component power law (2PL) fit to \Gaia~DR3 data in the $8-9\,\rm{kpc}$ radial bin and with $\vmin=310\,\rm{km\,s^{-1}}$. Outlier distribution parameters not shown as the outlier distribution is consistent with 0 in this bin.}
    \label{fig:corner2PL}
\end{figure*}

\begin{figure*}
    \centering
    \includegraphics[width=0.8\textwidth]{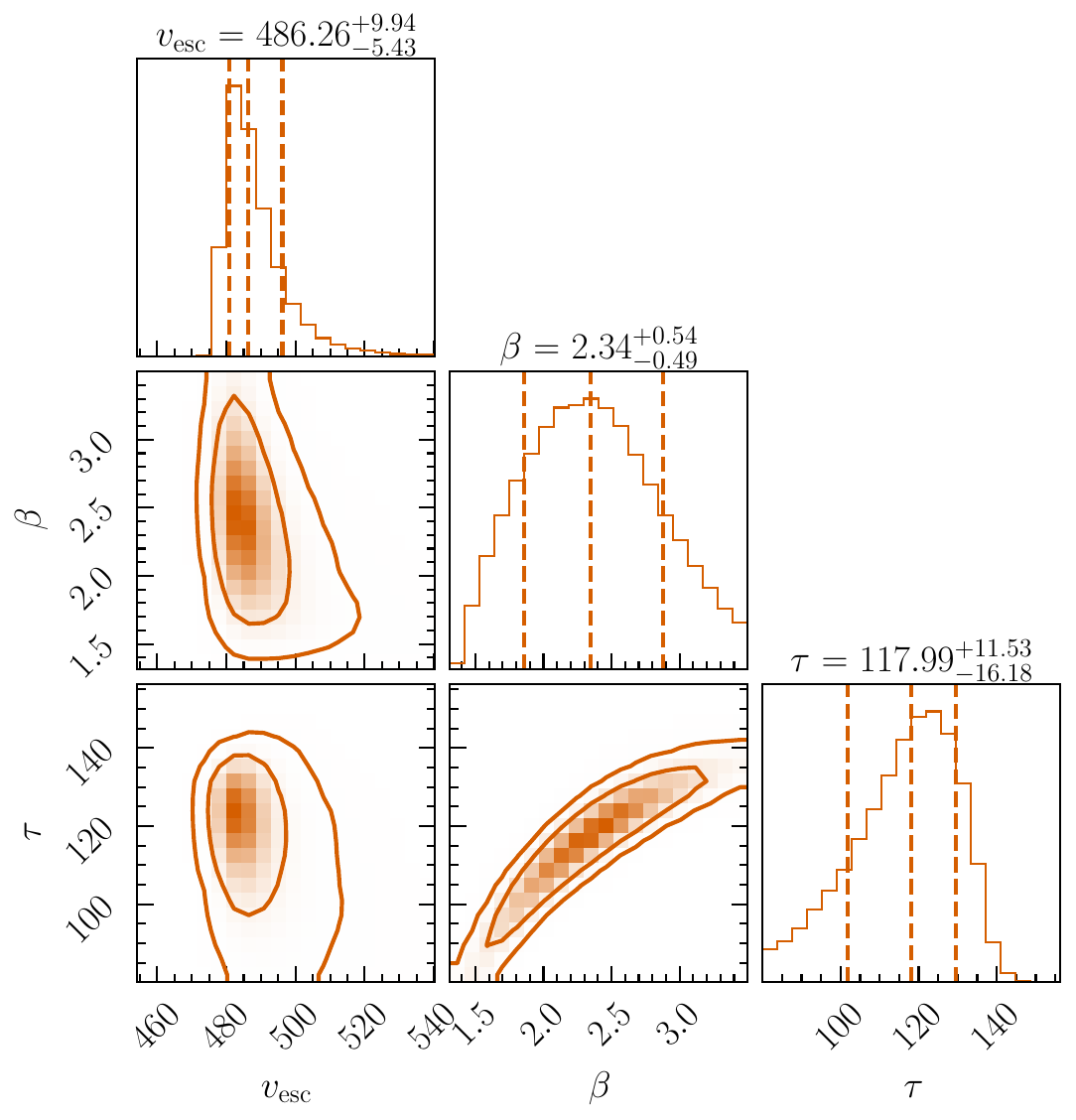}
    \caption{Stretched exponential power law (SEPL) fit to \Gaia~DR3 data in the $8-9\,\rm{kpc}$ radial bin and with $\vmin=310\,\rm{km\,s^{-1}}$. Outlier distribution parameters not shown as the outlier distribution is consistent with 0 in this bin.}
    \label{fig:cornerSEPL}
\end{figure*}

\begin{figure*}
  \centering
  \includegraphics[width=\textwidth]{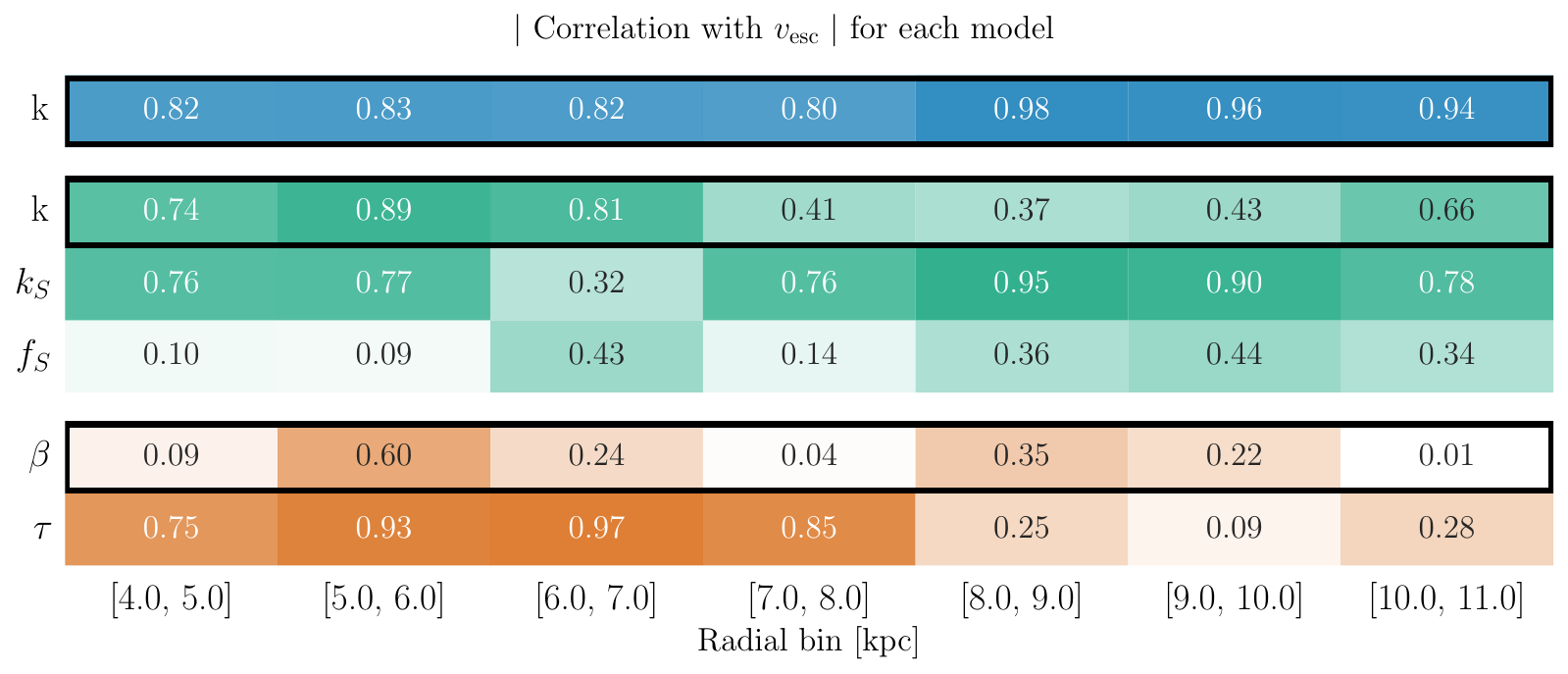}
  \caption{Correlations between $v_{\rm{esc}}$ and other model parameters. First row is the 1-component power law (1PL) correlations, next 3 rows are 2-component power law (2PL), and final two rows are correlations with the stretched exponential (SEPL) model parameters. Outlier model correlations not shown, as they exhibit no significant difference across models. Black outlines indicate parameters which are often not properly converged, and their correlations with the parameter of interest, $\vesc$.}
  \label{fig:correlation}
\end{figure*}

\end{document}